\documentclass[aps,prx,twocolumn,superscriptaddress,showpacs]{revtex4-1}
\usepackage{amsmath,amssymb,amsfonts,graphics,epsfig,epstopdf,color,verbatim,ulem,tabularx}
\usepackage[colorlinks,linkcolor=blue,citecolor=blue,urlcolor=blue]{hyperref}
\usepackage{wasysym}
\usepackage{color}
\newcommand{\ii}{\mathrm{i}}
\renewcommand{\Re}{\mathrm{Re}}
\renewcommand{\Im}{\mathrm{Im}}
\newcommand{\eqnref}[1]{Eq.\,\eqref{#1}}
\newcommand{\figref}[1]{Fig.\,\ref{#1}}

\newcommand{\beq}{\begin{equation}}
\newcommand{\eeq}{\end{equation}}
\newcommand{\beqn}{\begin{eqnarray}}
\newcommand{\eeqn}{\end{eqnarray}}
\graphicspath{{Figures/}}

\begin{document}

\title{Duality between the deconfined quantum-critical point and the \\ bosonic topological transition}

\author{Yan Qi Qin}
\affiliation{Institute of Physics, Chinese Academy of Sciences,
Beijing 100190, China} \affiliation{University of Chinese Academy
of Sciences, Beijing 100049, China}

\author{Yuan-Yao He}
\affiliation{Department of Physics, Renmin University of China, Beijing 100872, China}

\author{Yi-Zhuang You}
\affiliation{Department of Physics, Harvard University, Cambridge, MA 02138, USA}

\author{Zhong-Yi Lu}
\affiliation{Department of Physics, Renmin University of China, Beijing 100872, China}

\author{Arnab Sen}\affiliation{Department of Theoretical Physics, Indian Association for the Cultivation of Science, Jadavpur, Kolkata 700032, India}

\author{Anders W. Sandvik}
\affiliation{Department of Physics, Boston University, 590 Commonwealth Avenue, Boston, Massachusetts 02215, USA}

\author{Cenke Xu}
\affiliation{Department of physics, University of California, Santa Barbara, CA 93106, USA}

\author{Zi Yang Meng}
\affiliation{Institute of Physics, Chinese Academy of Sciences, Beijing 100190, China}
\affiliation{University of Chinese Academy of Sciences, Beijing 100049, China}

\date{\today}

\begin{abstract}
Recently significant progress has been made in $(2+1)$-dimensional
conformal field theories without supersymmetry. In particular, it
was realized that different Lagrangians may be related by hidden
dualities, i.e., seemingly different field theories may actually
be identical in the infrared limit. Among all the proposed
dualities, one has attracted particular interest in the field of
strongly-correlated quantum-matter systems: the one relating the
easy-plane noncompact CP$^1$ model (NCCP$^1$) and noncompact
quantum electrodynamics (QED) with two flavors ($N = 2$) of
massless two-component Dirac fermions. The easy-plane NCCP$^1$
model is the field theory of the putative deconfined
quantum-critical point separating a planar (XY) antiferromagnet
and a dimerized (valence-bond solid) ground state, while $N=2$
noncompact QED is the theory for the transition between a bosonic
symmetry-protected topological phase and a trivial Mott insulator.
In this work we present strong numerical support for the proposed
duality. We realize the $N=2$ noncompact QED at a critical point
of an interacting fermion model on the bilayer honeycomb lattice
and study it using determinant quantum Monte Carlo (QMC)
simulations. Using stochastic series expansion QMC, we study a
planar version of the $S=1/2$ $J$-$Q$ spin Hamiltonian (a quantum
XY-model with additional multi-spin couplings) and show that it
hosts a continuous transition between the XY magnet and the
valence-bond solid. The duality between the two systems, following
from a mapping of their phase diagrams extending from their
respective critical points, is supported by the good agreement
between the critical exponents according to the proposed duality
relationships.

\end{abstract}


\maketitle

\section{Introduction}

A duality in physics is an equivalence of different mathematical
descriptions of a system or a state of matter, established through
a mapping by change of variables. The simplest example is the
particle-wave duality in quantum mechanics, where the duality
transformation is a change of basis by Fourier transformation, and
the chosen basis dictates the variables used to describe the
system. In classical statistical mechanics, the most famous
duality is the Kramers-Wannier duality of the two-dimensional (2D)
Ising model \cite{kramers41}. Here the low- and high-temperature
expansions of the partition function can be related to each other
by identifying a one-to-one correspondence between the terms in
the two different series, thus establishing an exact mapping
between the ordered and disordered phases and the corresponding
collective variables. In this case the critical point is also a
self-duality point. In the 3D Ising model, one can instead find a
different model whose high-temperature expansion stands in a
direct one-to-one correspondence with the low-temperature
expansion of the the Ising gauge model~\cite{Kogut1979,savit}.
Many other examples of dualities have been established, e.g., the
well-known equivalence between the 3D O(2) Wilson-Fisher fixed
point and the 3D Higgs transition with a noncompact U(1) gauge
field~\cite{Peskin,halperindual,leedual}~\footnote{Here the term
''noncompact'' means that the U(1) gauge flux is a conserved
quantity.}.

In analogy with the Ising examples mentioned above, it is some
times possible to transform a quantum field theory at strong
coupling into an equivalent dual theory at weak coupling. The
untractable original problem can then be solved by means of
perturbative methods applied to the dual theory. Such strong-weak
duality (and the more general ``S-duality" form) was established
in certain supersymmetric Yang-Mills theories
\cite{Montonen1977,seiberg1995,seibergwitten1,seibergwitten2} and
Abelian gauge theory without
super-symmetry~\cite{cardy1,cardy2,wittenu1}. In 1D quantum
systems (i.e., in $1+1$ space-time dimensions) a well known
fermion-boson duality is achieved by bosonization of an
interacting fermion system through a non-local transformation
\cite{coleman,mandelstam,luther,wittenbosonization,Giamarchi2004}.
Usually, in the bosonized formalism interactions can be more
easily treated than in the original fermion model.

In cases where no formal mapping is known, two Lagrangians that
look different in the ultraviolet may still flow (under the
renormalization group) to the same theory in the infrared, i.e.,
these seemingly different field theories actually represent exactly
the same low-energy physics. Such a duality goes a step beyond
the more familiar concept of universality, by which systems
(models or real materials) with the same dimensionality and global
symmetries exhibit identical scaling properties at their classical
or quantum critical points. Such systems share the same effective
critical low-energy field-theory description. For example, the
critical points of the Bose-Hubbard model and the quantum rotor
model are in the same universality class. A duality transformation
usually changes the description of the system into a form based on
nonlocal objects or defects of the original description.
On a practical level, the existence of a dual
field theory means that there is a non-trivial choice of which
description to use, and one of them (and not necessarily the
originally most obvious one) may pose a more tractable setup for
calculations.

Even if no strong-weak transformation exists (or is known), a
difficult or non-tractable strong-coupling problem can some times
be shown to be dual to a different strong-coupling problem that is
tractable with some specific computational technique. In
particular, the dual problem may be more easily solvable using
powerful numerical (lattice) methods. In this paper we will
explore such a recently proposed duality between two different
$(2+1)$D Lagrangians that respectively involve fermionic and bosonic
matter fields coupled with a gauge field~\cite{potterdual,SO5}.
Both theories are of great current interest in the context of
strongly correlated electrons in two dimensions. Our aim here is
to identify a duality between the systems by establishing
corresponding lattice models realizing the two low-energy
theories.

\begin{figure*}[t]
\includegraphics[width=\textwidth]{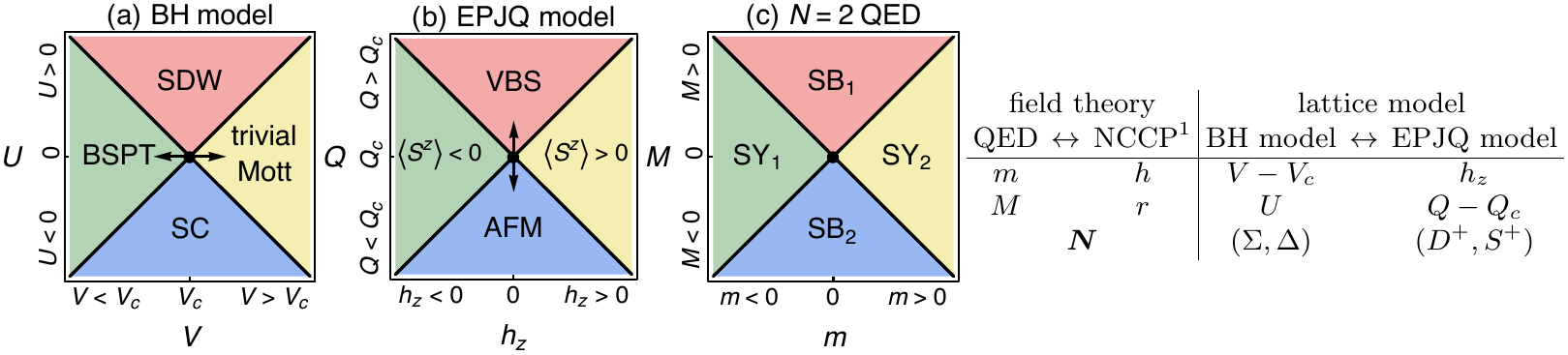}
\caption{Schematic phase diagrams of (a) the bilayer honeycomb
  (BH) model, (b) the easy-plane $J$-$Q_3$ (EPJQ) model, and (c) the $N=2$
  QED theory. In (a), the BH model contains two symmetry-breaking phases:
the spin-density-wave (SDW) and superconducting (SC) phases, and
two symmetric phases: the bosonic symmetry-protected topological
(BSPT) and the trivial Mott-insulating phases. In (b), the EPJQ
model also contains two symmetry-breaking phases: the N\'eel
antiferromagnetic (AFM) phase and the valence-bond solid (VBS)
phase, and two spin-polarized phases induced by an external
staggered field. In (c), as was shown in
Ref.~\cite{Tarun_PRB2013,lulee,SO5}, when tuning the two masses
$m$ and $M$, the $N=2$ QED theory also has two symmetry-breaking
(SB) phases and two symmetric (SY) phases, one of which is the
BSPT state. In all models, the four phases merge at the deconfined
quantum critical point. Phases of the same color can be mapped to
each other among the models, following the duality relations
proposed in the table on the right. The double arrows in (a) and
(b) indicate the quantum phase transitions investigated
numerically in this paper.} \label{fig:phasediagrams}
\end{figure*}

We follow the recent proposal that $(2+1)$D quantum
electrodynamics (QED) with noncompact gauge field and two flavors
of Dirac fermions is dual to the critical point of the easy-plane
NCCP$^1$ model (the bosonic QED with two flavors of complex
bosons)~\cite{potterdual,SO5}. On the lattice, we realize the
former with an interacting fermion model with spin-orbit coupling
on the bilayer honeycomb lattice (BH), which hosts a quantum phase
transition between a (bosonic) symmetry-protected topological
state and a trivial Mott
insulator~\cite{kevinQSH,mengQSH2,Wu2016}. It was proposed that
this transition is described by $N=2$ noncompact
QED~\cite{groverashvin,lulee}. To realize the low-energy physics
of the NCCP$^1$ theory, in this paper we introduce a planar
variant of the spin $S=1/2$ $J$-$Q$ Hamiltonian (a Heisenberg
model with additional multi-spin couplings \cite{Sandvik2007}),
dubbed the easy-plane $J$-$Q$ (EPJQ) model, and show that it hosts
a deconfined quantum-critical point
\cite{deconfine1,deconfine2,Senthil2006} separating
antiferromagnetic (AFM) and dimerized (valence-bond-solid, VBS)
ground states (similar to the case of the $J$-$Q$ model with full
spin SU(2) symmetry~\cite{Shao2016}, but with different
universality due to the lowered symmetry). Our numerical (quantum
Monte Carlo, QMC) results establishes the critical-point
universality and duality between the phase diagrams of the two
models. With the EPJQ model being much easier to study on large
scales with QMC simulations than the fermionic model, the duality
that we establish here allows for detailed studies of the
topological transition of the latter, through the analogue of the
deconfined quantum-critical point. The phase diagrams and
dualities studied are summarized in Fig.~\ref{fig:phasediagrams}.

The rest of the paper is organized as follows: In
Sec.~\ref{sec:fieldtheories} we present the details of the two
field theories and their putative duality, and in
Sec.~\ref{sec:latticemodels} we define the lattice models and the
proposed mappings relating their phase diagrams to each other. In
Sec.~\ref{sec:numericaljq} and Sec.~\ref{sec:numericalbh} we
present the numerical results for the EPJQ and BH models,
respectively. We conclude with a brief summary and discussion of
the results in Sec.~\ref{sec:discussion}. In Appendix \ref{app:crossingpoint}
we present further technical details on the analysis of the critical
exponents, and in Appendix  \ref{app:planarJQ} we compare results for
different variants of the EPJQ model (with different degrees of
spin-anisotropy).

\section{Continuum field theories}
\label{sec:fieldtheories}

The bosonic particle-vortex duality mentioned in the introduction
was recently generalized to a model with fermionic matter
\cite{son,wangsenthil0,wangsenthil1,wangsenthil2,maxashvin,seiberg1,kachru1,kachru2,mrossduality},
in the form of a $(2+1)$D QED Lagrangian with a single flavor of a
two component Dirac fermion and noncompact gauge field, i.e.,
$N=1$ QED. This theory is dual to that of a noninteracting Dirac
fermion in the infrared limit~\footnote{A more precise form of the
dual $(2+1)$D QED is given in Ref.~\onlinecite{seiberg1}.}. Based
on this $N=1$ duality, Ref.~\cite{xudual} showed that $(2+1)$D QED
with noncompact gauge field and $N=2$ flavors of Dirac fermions is
self-dual. This is also a fermionic version of the self-duality of
the easy-plane NCCP$^{1}$ model (which can be regarded as $N=2$
bosonic QED)~\cite{ashvinlesik,deconfine1,deconfine2}. The
self-duality of the $N=2$ QED Lagrangian was also verified with
different derivations~\cite{mrossduality,karchtong,seiberg2}.
Unlike the case of $N=1$, there is no equivalent noninteracting
description of $(2+1)$D QED with $N=2$, however. Because of its
self-duality, $N=2$ QED hosts an (emergent) O($4$) symmetry in the
infrared, which factorizes into the two independent SU($2$) flavor
symmetries on the two sides of the self-dual point.

More recently, based on the previous fermion-boson
duality~\cite{seiberg1}, it was argued that $N=2$ QED is also dual
to the easy-plane NCCP$^1$ model at the critical
point~\cite{potterdual,SO5}. These two field theories can be
written as
\begin{subequations}
\label{duality}
\begin{align}
& \mathcal{L}_\text{QED} =
\bar{\psi} \gamma\cdot (\partial - \ii a) \psi + m \bar{\psi}\psi + M \bar{\psi}\sigma^3 \psi \label{duality1}, \\
& \mathcal{L}_\text{CP$^1$} =
|(\partial - \ii b)z|^2 + g |z|^4 + r z^\dagger z + h z^\dagger \sigma^3 z \label{duality2},
\end{align}
\end{subequations}
where $\psi$ and $z$ are two-component Dirac fermion and complex
boson fields coupled to non-compact U(1) gauge fields, $a$ and
$b$, respectively. The duality maps the variables $(m,M)$ to
$(h,r)$. Moreover, both theories in \eqnref{duality} are
individually self-dual. The putative duality between the two
theories implies that the easy-plane NCCP$^1$ model should also
have an emergent O($4$) symmetry at its critical point, which is
not immediately obvious in \eqnref{duality2}. The corresponding
O($4$) order parameter is \beqn \boldsymbol{N} = \left( z^\dagger \sigma^x z,
z^\dagger \sigma^y z, \mathrm{Re}[\mathcal{M}_b],
\mathrm{Im}[\mathcal{M}_b] \right), \label{vector} \eeqn where
$\mathcal{M}_b$ is the monopole operator (gauge flux annihilation
operator) of the gauge field $b$.

The proposed duality between the two theories in \eqnref{duality}
leads to very strong predictions for relationships between their
properties. For example, the scaling dimension of $m$ in $N = 2$
QED should be precisely the same as the scaling dimension of $h$
in the easy-plane NCCP$^1$ at its critical point $r=0$, while the
scaling dimension of $M$ should be the same as that of $r$. Also,
as a  consequence of these dualities, i.e., the emergent O($4$)
symmetry of the two theories, the four components of $\boldsymbol{N}$
should all have the same scaling dimension at the critical point.

Although the duality can be observed and ``derived'' based on
various arguments, it has not been rigorously proven yet. Both
\eqnref{duality1} and \eqnref{duality2} are strongly interacting
conformal field theories, and there is no obvious analytical
method that can provide rigorous results for either case. However,
both theories can presumably be realized using lattice models,
which can be simulated using numerical methods. The goal of this
work is to compare the quantitative properties of such lattice
models and look for evidence of the proposed duality. As we will
show in the later sections, within small error bars of the critical
exponents obtained using QMC simulations, our results confirm
several predictions of the duality.

Because the duality of the two theories in \eqnref{duality} was
derived based on the assumption of the basic fermion-boson
duality~\cite{wufisher,seiberg1}, a proof of the former duality
indirectly also proves the latter. In principle this result can
lead to a number of further dualities between different fermionic
and bosonic Lagrangians. Thus, the impact of our work is not
limited to the proof of the duality between \eqnref{duality1} and
\eqnref{duality2}, but also provides justification for many other
cases.

\section{Lattice models}
\label{sec:latticemodels}

The easy-plane NCCP$^1$ model is the field theory that presumably
describes the deconfined quantum-critical point between an
in-plane (XY) antiferromagnet (AFM) and a valence-bond solid (VBS)
\cite{ashvinlesik,deconfine1,deconfine2}. This transition in the
case of full SU(2) symmetry of the Hamiltonian has been realized
by the $J$-$Q$ and related models, and these have been extensively
simulated numerically using unbiased QMC techniques
\cite{Sandvik2002,Sandvik2007,Lou2009,Sen2010,Sandvik2010,Nahum2011,Harada2013,Pujar2013,Pujari2015,Nahum2015a,Nahum2015b,Shao2016}.
Although there are studies that indicate that some version of the
$J$-$Q$ model with an inplane spin symmetry and other U(1)
symmetric models should lead to a first order transition
\cite{Kuklov08,Geraedts2012,Jonathan2016,Jonathan2017}, in this work we demonstrate that a different
model, the EPJQ model, instead leads to a continuous transition in
some regions of its parameter space. The $r$ and $h$ terms in
\eqnref{duality} correspond to the distance from the critical
point, $Q - Q_c$, and the staggered magnetic field $h_z (-)^i
S^z_i$, respectively, in the lattice model. The components of the
O(4) vector in \eqnref{vector} correspond to the two-component
easy-plane N\'eel and two-component VBS (dimer) order parameters
of the EPJQ model.

The $N=2$ QED action has been simulated directly using a lattice
QED model~\cite{Karthik2016}, and the scaling dimension of $M$ was
computed in this way. Also, $N = 2$ QED with a noncompact U(1)
gauge field is the effective theory that describes the transition
between the bosonic symmetry protected topological (BSPT) state
and a trivial Mott state in $2D$ \cite{groverashvin,lulee}. This
transition was also realized in an interacting fermion model on a
bilayer honeycomb (BH) lattice introduced in
Refs.~\cite{kevinQSH,mengQSH2} and simulated
\cite{kevinQSH,mengQSH2,He2016Topinvb,Wu2016} with a determinantal
QMC method (DQMC)~\cite{Meng2010,Meng2014}. The $m$ and $M$ terms
in \eqnref{duality} correspond to two different interactions in
the lattice model, namely, the interlayer pair-hopping $V-V_c$,
measured with respect to its critical value, and the Hubbard-like
on-site interaction $U$; the Hamiltonian will be specified in
detail below. The lattice model of Ref.~\cite{mengQSH2} has an
exact SO(4) symmetry that precisely corresponds to
the proposed emergent symmetry of the $N=2$ QED. It should
further be noted that the fermions in the BH model do not directly
correspond to the Dirac fermions of the $N=2$ QED action, because
the former are not coupled to any dynamical gauge field. The
relation between the two systems instead arises from the
correspondence of the gauge invariant fields of $N = 2$ QED to the
low-energy bosonic excitations of the BH model.

Using the BH model and the EPJQ model, the duality between the
$N=2$ QED and the NCCP$^1$ field theories can be realized on the
lattice. The O(4) vector $\boldsymbol{N}$ can also be conveniently
defined in both lattice models, with explicit forms that will be
explained below. Thus, the two systems can be investigated and
compared via unbiased large-scale QMC simulations---the pursuit
and achievement of this work.

In the following, we will first define the microscopic lattice
models in detail. In the subsequent sections \ref{sec:numericaljq}
and \ref{sec:numericalbh} we present comparative numerical studies of
the models and demonstrate strong support for the duality relations
listed in the table in Fig.~\ref{fig:phasediagrams}.

\subsection{Bilayer honeycomb model}
\label{sec:BHmodel}

The BH model is a fermionic model defined on a honeycomb
lattice \cite{kevinQSH,mengQSH2,He2016Topinvb,Wu2016}. On each
site, we define four flavors of fermions (two layers $\times$ two
spins);
\begin{equation}
c_{i}=(c_{i1\uparrow},c_{i1\downarrow},c_{i2\uparrow},c_{i2\downarrow})^\intercal.
\end{equation}
The Hamiltonian is
\begin{equation}
\label{eq:BH}
H_\text{BH}=H_\text{band}+H_\text{int},
\end{equation}
where the band and interaction terms are given by
\begin{subequations}
\begin{align}
  H_\text{band}&=-t\sum_{\langle ij\rangle}c_{i}^\dagger c_{j}+\lambda\sum_{\langle\!\langle i j\rangle\!\rangle}
  \ii \nu_{ij} (c_{i}^\dagger \sigma^3 c_{j}+\text{h.c.}),\\
  H_\text{int}&=V\sum_{i}(c_{i1\uparrow}^\dagger c_{i2\uparrow}
   c_{i1\downarrow}^\dagger c_{i2\downarrow}+\text{h.c.}),
\end{align}
\end{subequations}
where $\langle i j\rangle$ and $\langle\!\langle i
j\rangle\!\rangle$ denote nearest-neighbor intra- and inter-layer
site pairs, respectively. The band Hamiltonian $H_\text{band}$ is
just two copies of the Kane-Mele model~\cite{kane2005a}, which
drives the fermion into a quantum spin Hall state with spin Hall
conductance $\sigma_\text{sH}=\pm2$ (depending on the sign of the
spin-orbit coupling $\lambda$). Including a weak interaction $V$,
the bilayer quantum spin Hall state automatically becomes a BSPT
state~\cite{mengQSH2,Wu2016,youspn}, where only the bosonic O(4)
vector $\boldsymbol{N}$ remains gapless (and protected) at the
edge, while the fermionic excitations are gapped out (as will be
discussed in more detail below). However, a strong interlayer
pair-hopping interaction $V$ eventually favors a direct product
state of anti-bonding Cooper pairs. In the strong interaction
limit ($V\to\infty$), the ground state of the BH model reads
\begin{equation}\label{eq:gs}
|\text{GS}\rangle=\prod_{i}(c_{i1\uparrow}^\dagger
c_{i1\downarrow}^\dagger-c_{i2\uparrow}^\dagger
c_{i2\downarrow}^\dagger)|0_c\rangle,
\end{equation}
with $|0_c\rangle$ being the fermion vacuum state. This state has
no quantum spin Hall conductance, i.e., $\sigma_\text{sH}=0$, and,
more importantly, it is a direct product of local wave
functions, hence dubbed trivial Mott insulator state. It was found
numerically that there is a direct continuous transition between
the BSPT and the trivial Mott phases at
$V_c/t=2.82(1)$~\cite{mengQSH2,He2016Topinvb}, where the
single-particle excitation gap does not close but the excitation
gap associated with the bosonic O(4) vector closes and the
quantized spin Hall conductance changes from $\pm2$ to $0$.

The low-energy bosonic fluctuations around the critical point form
an O(4) vector, with $\boldsymbol{N}=(\Re\Sigma,\Im\Sigma,\Re\Delta,\Im\Delta)$ and
the components are
\begin{subequations}
\begin{align}
\Sigma_i&=(-1)^i(c_{i1\uparrow}^\dagger c_{i2\downarrow}+c_{i2\uparrow}^\dagger c_{i1\downarrow}), \label{eq:O4vector1}\\
\Delta_i&=(c_{i1\downarrow}c_{i1\uparrow}-c_{i2\downarrow}c_{i2\uparrow}), \label{eq:O4vector2}
\end{align}
\label{eq:O4vector}
\end{subequations}
\noindent where $\Sigma$ carries spin and $\Delta$ carries charge.
The BH model \eqnref{eq:BH} respects the global $\text{SO}(4)$
symmetry of the vector $\boldsymbol{N}$. If the symmetry is
lowered to
$\text{U}(1)_\text{spin}\times\text{U}(1)_\text{charge}$, then,
based on the analysis of $N=2$ QED, in principle the mass term $M
\bar{\psi} \sigma^3 \psi$ is allowed; hence the BSPT-Mott
transition is unstable towards spontaneous symmetry-breaking of
the remaining symmetries. The symmetry of the mass term $M
\bar{\psi} \sigma^3 \psi$ is identical to the following
Hubbard-like interaction (both forming a $(1,1)$ representation of
the SO(4)):
\begin{equation}
\frac{U}{2}\sum_i(\Delta_i^\dagger\Delta_i+\Delta_i\Delta_i^\dagger-\Sigma_i^\dagger\Sigma_i-\Sigma_i\Sigma_i^\dagger)=U\sum_{i}\rho_{i\uparrow}\rho_{i\downarrow}.
\end{equation}
Here $\rho_{i\sigma}$ is the density operator (for $\sigma=\uparrow,\downarrow$ spins),
\begin{equation}
\rho_{i\sigma}=(c_{i1\sigma}^\dagger c_{i1\sigma}+c_{i2\sigma}^\dagger c_{i2\sigma}-1),
\end{equation}
which counts the number of $\sigma$-spin fermions in both layers on site $i$ with
respect to half-filling. The repulsive $U>0$ (or attractive $U<0$) interaction drives spin
$\langle\Sigma\rangle\neq0$ (or charge $\langle\Delta\rangle\neq0$) condensation, leading to
a spin-density wave (SDW)~\cite{Wu2016} (or superconducting) phase that breaks the
$\text{U}(1)_\text{spin}$ [or $\text{U}(1)_\text{charge}$] symmetry spontaneously. This process
is illustrated in the schematic phase diagram Fig.~\ref{fig:phasediagrams}(a).

\subsection{Easy-plane JQ model}
\label{sec:JQmodel}

Our EPJQ model is a spin-$1/2$ system with anisotropic antiferromagnetic couplings which we here define
on the simple square lattice of $L^2$ sites and periodic boundary conditions. It is a ``cousin'' model of
the previously studied SU(2)$_\text{spin}$ $J$-$Q_3$ model \cite{Lou2009,Sen2010,Sandvik2010}which
in turn is an extension of the original $J$-$Q$, or $J$-$Q_2$, model \cite{Sandvik2007}. Starting from
the spin-$1/2$ operator $\boldsymbol{S}_{i}$ on each site $i$, we define the singlet-projection operator
on lattice link $ij$;
\begin{equation}
P_{ij}=\tfrac{1}{4}-\boldsymbol{S}_i\cdot\boldsymbol{S}_j,
\end{equation}
then the model Hamiltonian reads
\begin{equation}
H_\text{JQ}=-J\sum_{\langle ij\rangle}(P_{ij}-\Delta
S_i^zS_j^z)-Q\hskip-2mm\sum_{\langle ijklmn \rangle}\hskip-2mm P_{ij}P_{kl}P_{mn},
\label{eq:JQmodel}
\end{equation}
where the $\Delta S_i^zS_j^z$ term for $\Delta \in (0,1]$ introduces the easy-plane anisotropy that breaks the $\text{SU}(2)_\text{spin}$
symmetry down to $\text{U}(1)_\text{spin}$ explicitly. We have studied two cases: the maximally-planar case $\Delta=1$ and the less
extreme case $\Delta=1/2$. In the latter case, we observe very good scaling behaviors indicating a continuous transition, with rapidly
decaying (with the system size $L$) scaling corrections, while for $\Delta=1$ the behavior suggests a first-order transition. Thus, the model
may harbor a tricritical point separating first-order and continuous transitions somewhere between $\Delta=1/2$ and $\Delta=1$. However, we
will leave the possible tricritical point to future investigation. As far as the duality is concerned, in this section we
discuss our results for $\Delta=1/2$, and in Appendix~\ref{app:planarJQ} we present results for $\Delta=1$.

We set $J+Q=1$ in the simulations and define the control parameter as the ratio
\begin{equation}
q=\frac{Q}{J+Q}.
\end{equation}
For small $q$, the model essentially reduces to an XXZ model, which has an XYAFM ground state that breaks the $\text{U}(1)_\text{spin}$
symmetry spontaneously. When $q$ is large, the dimer interaction favors a VBS (columnar-dimerized) ground state, which breaks the lattice
$C_4$ rotation symmetry as in the $\text{SU}(2)_{\text{spin}}$ $J$-$Q_2$ and $J$-$Q_3$ models
\cite{Sandvik2002,Sandvik2007,Lou2009,Sen2010,Sandvik2010,Shao2016}, where previous QMC studies found a direct continuous AFM-VBS transition.
Here we demonstrate that the continuous transition persists in EPJQ model, \eqnref{eq:JQmodel}, with $\Delta=1/2$.
The reason for choosing the $Q_3$ term (three-dimer interaction) instead of the simpler $Q_2$ interaction (two-dimer coupling) is that
it produces a more robust VBS order when the ratio $q$ is large, thus leading to a smaller critical-point value [as in the SU(2) case
\cite{Lou2009,Sen2010}] with more clearly observable flows to the VBS state on that side of the transition.

The XYAFM order in the EPJQ model can be directly probed by the local spin components
$S^x$ and $S^y$, and we will also study the critical fluctuations in the $S^z$ component.
We will often not write out the staggered phase factor $(-1)^{x_i+y_i}$ corresponding to AFM
order explicitly (here $x_i$ and $y_i$ refer to the integer-valued lattice coordinates of site $i$);
in fact in the case of the XY-anisotropy in a model with bipartite interactions, the phase can also simply
be transformed away with a sublattice rotation (and then the XYAFM phase maps directly onto hard-core bosons in the
superfluid state). In our simulations the staggered phase is absent for the $S^x$ and $S^y$ components but present
for the $S^z$ component. The AFM-VBS transition is unstable towards an axial Zeeman field, when
$H_{\rm JQ} \to H_{\rm JQ} + H_{\rm Z}$ with
\begin{equation}
H_{\rm Z} = -h_z\sum_{i}(-1)^{x_i+y_i}S_i^z,
\end{equation}
which drives the system to the spin-polarized phase with $\langle S^z\rangle\neq 0$, as illustrated in the
schematic phase diagram in Fig.~\ref{fig:phasediagrams}(b). Here we consider only $h=0$.

To study the columnar VBS (dimer) order realized in the EPJQ model, we define
\begin{subequations}
  \label{dops}
  \begin{align}
  & D^x_{i}=(-1)^{x_i}{\bf S}_i \cdot {\bf S}_{i+\hat x},\\
  & D^y_{i}=(-1)^{y_i}{\bf S}_i \cdot {\bf S}_{i+\hat y},
  \end{align}
\end{subequations}
where ${i+\hat x}$ and ${i+\hat y}$ denote neighbors of site $i$ in the positive $x$ and $y$-direction,
respectively. At the critical point, the proposed self-duality (through the putative duality with $N=2$ QED) implies
that the $C_4$ rotation symmetry and the $\text{U}(1)_\text{spin}$ symmetry are enlarged into an emergent $\text{O}(4)$
symmetry, such that the components of the $\text{O}(4)$ vector (after some proper normalization)
\begin{equation}
\boldsymbol{N}=(D^x,D^y,S^x,S^y),
\end{equation}
should all have the same scaling dimension~\cite{SO5}.

\subsection{Duality relations}

Fig.~\ref{fig:phasediagrams} summarizes the intuitive duality relations among the BH, EPJQ, and QED models;
this can also be observed from the similarity of their four-quadrant phase diagrams~\cite{You:2016fv}. To numerically
prove the validity of these duality relations, in this work we investigate the following critical behaviors at
the BSPT--Mott transition in the BH model:
\begin{subequations}
\label{eq:BHexponents}
\begin{align}
\xi&\sim|V-V_c|^{-\nu_\text{BH}}, \label{eq:BHexponents1}\\
\langle\rho_{i\uparrow}\rho_{i\downarrow}\rho_{j\uparrow}\rho_{j\downarrow}\rangle&\sim|\boldsymbol{r}_{ij}|^{-1-\eta^{\rho}_{\text{BH}}}, \label{eq:BHexponents2} \\
\langle\Delta_{i}^\dagger \Delta_j\rangle&\sim|\boldsymbol{r}_{ij}|^{-1-\eta^{\Delta}_{\text{BH}}}; \label{eq:BHexponents3}
\end{align}
\end{subequations}
where $\boldsymbol{r}_{ij}$ is the lattice vector separating the sites $i,j$, $\xi$ is the correlation length of the critical O(4) bosonic
modes of the system, and the density $\rho_{i\sigma}$ and pairing $\Delta_{i}$ operators have been defined in Sec.~\ref{sec:BHmodel}. We also
study the following expected critical scaling behavior at the AFM--VBS transition in the EPJQ model:
\begin{subequations}\label{eq:JQexponents}
\begin{align}
\xi&\sim|Q-Q_c|^{-\nu^{xy}_\text{JQ}},\label{eq:JQexponents1} \\
\langle S_i^zS_j^z\rangle&\sim|\boldsymbol{r}_{ij}|^{-1-\eta^{z}_{\text{JQ}}},\label{eq:JQexponents2}\\
\langle S_i^+S_j^-\rangle&\sim|\boldsymbol{r}_{ij}|^{-1-\eta^{xy}_{\text{JQ}}}. \label{eq:JQexponents3}
\end{align}
\end{subequations}
where $\xi$ is the correlation length of the easy-plane spins.

If the duality in \eqnref{duality} is correct, and provided that $N=2$ QED is indeed the theory for the BSPT--Mott transition,
then the exponents defined above must satisfy the following relationships \cite{SO5}:
\begin{subequations}
\begin{align}
&3-\frac{1}{\nu_\text{BH}}=\frac{1+ \eta^z_\text{JQ}}{2},
\label{eq:dualityrelations1}\\
&3-\frac{1}{\nu^{xy}_\text{JQ}}=\frac{1+\eta_\text{QED}}{2} = \frac{1+\eta^{\rho}_{\text{BH}}}{2},
\label{eq:dualityrelations2}\\
&\eta^{\Delta}_\text{BH}=\eta^{xy}_\text{JQ}.
\label{eq:dualityrelations3}
\end{align}
\label{eq:dualityrelations}
\end{subequations}
Here $\eta_\text{QED}$ is the anomalous dimension of the fermion mass $\bar{\psi}\sigma^3\psi$, i.e., the $M$ mass term in our notation
in Eq.~(\ref{duality1}), which was numerically estimated in the recent lattice QED calculations in Ref.~\cite{Karthik2016}.

\section{Results for the EPJQ model}
\label{sec:numericaljq}

In this section we present our numerical results at the continuous AFM--VBS phase transition of the EPJQ model, obtained
using large-scale SSE-QMC~\cite{Sandvik1999,Syljuasen2002} simulations. Here we discuss only the case $\Delta=1/2$ in the Hamiltonian
\eqnref{eq:JQmodel}; some results for $\Delta=1$ are presented in Appendix \ref{app:planarJQ}. In the SSE simulations we
scaled the inverse temperature as $\beta=2L$, corresponding to the dynamic exponent $z=1$ ($\beta \sim L^z$) and staying
in the regime where the system is close to its ground state for each $L$. We consider $L$ up to $44$.

\subsection{Crossing-point analysis}
\label{sec:JQ_crossingpoint}

The first step is to determine the order of the transition and the
position of the critical point (if the transition is continuous).
To this end, following the recent example in
Ref.~\cite{Shao2016} for the SU(2) $J$-$Q$ model, we first
analyze crossing points of finite-size Binder cumulants, defined
for the AFM order parameter as
\begin{equation}
U(q,L)=2-\frac{\langle M^{4}_{xy}\rangle}{\langle M^{2}_{xy}\rangle^{2}},
\end{equation}
where $M^{2}_{xy}$ is the square of the easy-plane magnetization operator,
\begin{equation}
M^{2}_{xy}=\frac{1}{L^4}\sum_{i,j}(-1)^{i+j}S^{+}_{i}S^{-}_{j},
\label{eq:mxydef}
\end{equation}
and $M^{4}_{xy}$ is its square. The ``phenomenological renormalization'' underlying
the crossing-point analysis and our technical implementations of it are discussed
in Appendix \ref{app:crossingpoint}. Here we show our numerical results and analyze
them within the scaling relationships presented in the appendix.

\begin{figure}[t]
\begin{center}
\includegraphics[width=7.5cm]{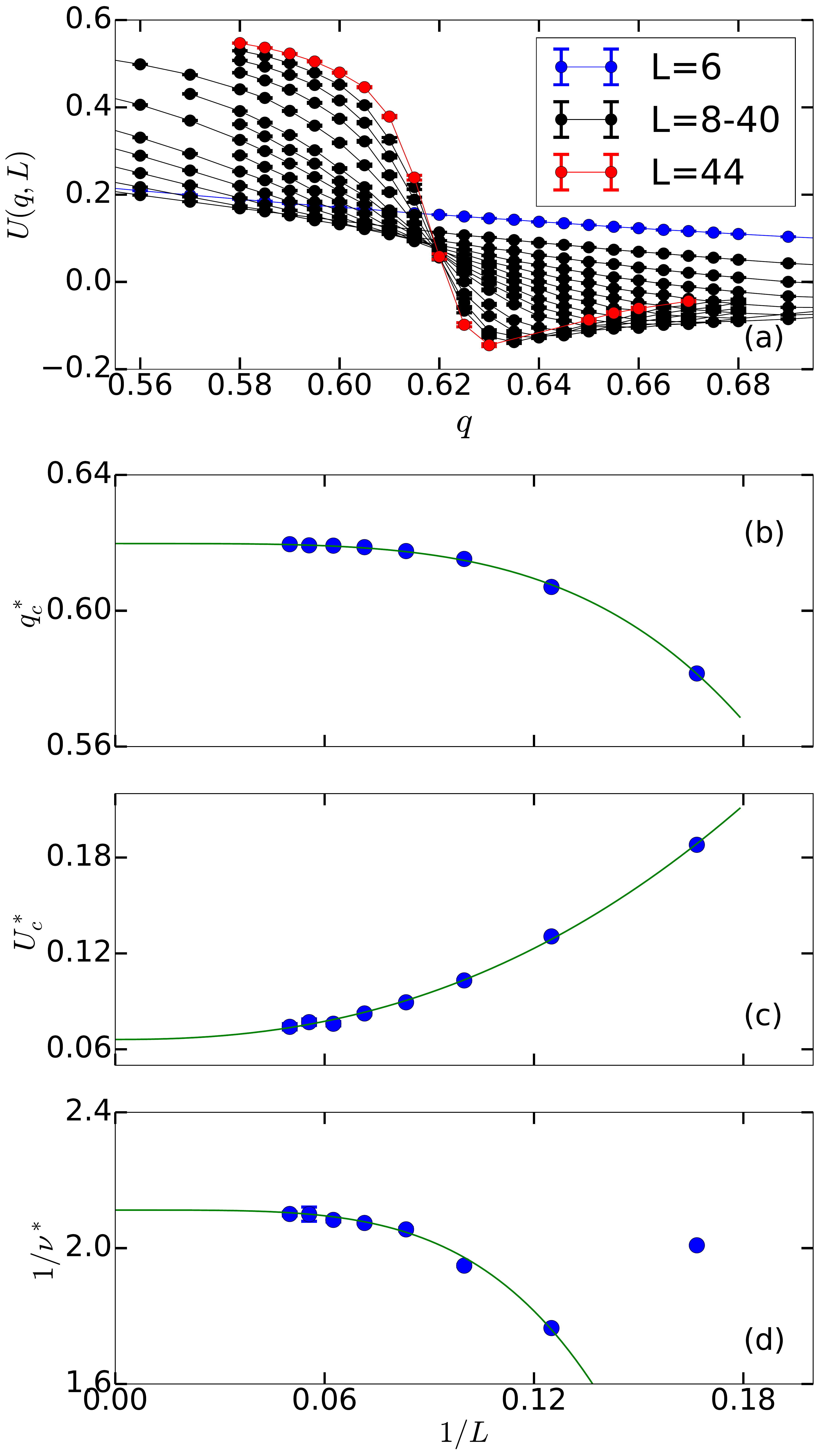}
\caption{Crossing-point analysis of the EPJQ model at $\Delta=1/2$. (a) $U(q,L)$ vs $q$ in the neighborhood of $q_c$ for several system sizes $L$.
  (b) The crossing points seen in (a) for system-size pairs $(L,2L)$, analyzed according to the expected finite-size scaling form,
  Eq.~(\ref{eq:qstarb}). The procedure including error analysis gives $q_c=0.6197(2)$ and $1/\nu^{xy}_{\text{JQ}}+\omega=4.0(2)$. (c) A similar analysis
  of $U^{*}_{c}(L)$, giving $\omega=2.3(1)$. (d) Finite-size estimates $\nu^{xy,*}_{\text{JQ}}$ of the correlation-length exponent defined in
  Eq.~(\ref{nustar}), using the slopes of the cumulants at the $(L,2L)$ crossing points. Analysis according to Eq.~(\ref{nustartscale}) gives
  $\nu^{xy}_{\text{JQ}}=0.48(2)$ for the exponent in the thermodynamic limit.} \label{fig:binder_crossing}
\end{center}
\end{figure}

As shown in \figref{fig:binder_crossing}(a), curves of $U(q,L)$ graphed for different $L$ cross each other at points tending to a
value $q_c$. In a finite-size system the deviation of $U(q,L)$ from the asymptotic crossing point depends on $L$ in a way that involves
a scaling-correction exponent. For a finite-size pair $(L,2L)$, the crossing is at $[q^{*}_c(L),U^{*}_{c}(L)]$ and at a continuous
transition one expects
\begin{equation}
q^{*}_{c}(L)=q_c+aL^{-(1/\nu^{xy}_{\text{JQ}}+\omega)},
\label{eq:qstarb}
\end{equation}
where $\nu^{xy}_\text{JQ}$ is the correlation-length exponent and $\omega$ is the smallest subleading exponent (which normally arises from
the leading irrelevant field). As shown in \figref{fig:binder_crossing}(b), an extrapolation with the above form to infinite size gives
$q_c=0.6197(2)$ (where the number in parenthesis indicates the one-standard-deviation error in the preceding digits, as obtained using numerical
error propagation with normal-distributed noise added to the data points) and $1/\nu^{xy}_\text{JQ} + \omega=4.0(2)$. The finite-size crossing
value of the cumulant itself, $U^{*}_{c}(L)$, should approach its thermodynamic limit $U_{c}$ as
\begin{equation}
U^{*}_{c}(L)=U_{c}+aL^{-\omega}.
\label{eq:qstarc}
\end{equation}
This form is used in Fig.~\ref{fig:binder_crossing}(c) and delivers $\omega=2.3(1)$. In principle we can now extract $\nu^{xy}_\text{JQ}$,
though the so obtained value and the independently determined value of $\omega$ in general should be viewed with some skepticism. The
exponents are often affected by neglected higher-order scaling corrections and should be regarded as an ``effective'' critical exponents
that flow to their correct values upon increasing the system sizes. Nevertheless, the fits with a single correction exponent are very good and
this may indicate that the next-order corrections are small.

The correlation length exponent $\nu^{xy}_\text{JQ}$ can be independently and more reliably obtained from the
slope of the Binder cumulant as
\begin{equation}
\label{nustar}
  \frac{1}{\nu^{xy,*}_{\text{JQ}}(L)}=\ln\Big(\frac{U'_{2}(2L)}{U'_{2}(L)}\Big)\frac{1}{\ln(2)},
\end{equation}
where $U'_{2}(L)$ is the derivative of $U(q,L)$ over $q$ evaluated at the crossing point between the $L$ and $2L$ curves (which we extract
by interpolating data close to the crossing point by cubic polynomials). The correlation-length exponent in
the thermodynamic limit can be extracted from the expected leading finite-size form
\begin{equation}
\label{nustartscale}
\frac{1}{\nu^{xy,*}_{\text{JQ}}(L)}=\frac{1}{\nu^{xy}_{\text{JQ}}}+L^{-\omega}.
\end{equation}
As shown in Fig.~\ref{fig:binder_crossing}(d), the fit to this form is statistically good if the smallest system sizes are excluded, and an extrapolation
then gives $\nu^{xy}_{\text{JQ}}=0.48(2)$. Thus, the combination $\nu^{xy}_{\text{JQ}} + \omega$ based on the independently evaluated two exponents
is in remarkably good agreement with the value of the sum extracted directly using Eq.~(\ref{eq:qstarb}) with the data in Fig.~\ref{fig:binder_crossing}(b).
This serves as a good consistency check and again indicates that the higher-order finite-size scaling corrections should be small (i.e., the following correction
exponents beyond $\omega$ must either have much larger values or the prefactors must be small, or both). Further support for this scenario can be observed
in Fig.~\ref{fig:binder_crossing}(d), where the data point for the smallest system size shown has a very large deviation from the good fit to the other
points, suggesting a very rapidly decaying correction.

In Fig.~\ref{fig:binder_crossing}(a) one may worry about the fact that the Binder cumulant forms a minimum extending to negative values as
the system size increases. A negative Binder cumulant often is taken as a sign of a first-order transition. However, it is now understood that
also some continuous transitions are associated with a negative Binder cumulant in the neighborhood of the critical point, reflecting non-universal
anomalies in the order-parameter distribution. Often the negative peak value grows slowly, e.g., logarithmically, with the system size, instead
of the much faster volume proportionality expected at a first-order phase transition. This issue is discussed with examples from classical
systems in Ref.~\cite{Jin2012}. Here we do not see any evidence of a fast divergence of the peak value; thus the transition should still
be continuous.

\subsection{Anomalous dimensions}
\label{sec:JQ_anomalous}

\begin{figure}[tp!]
\begin{center}
\includegraphics[width=80mm]{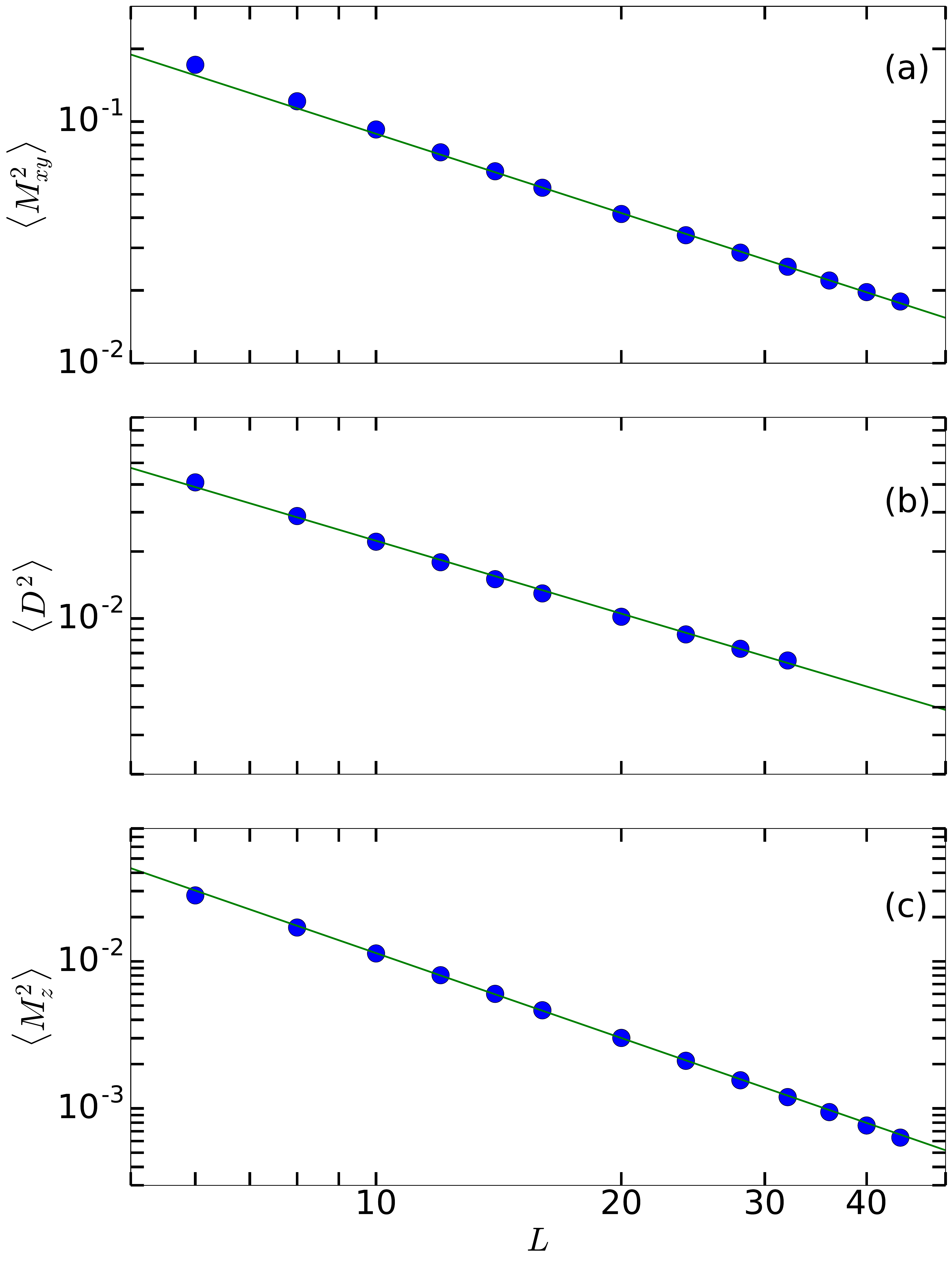}
\caption{Extraction of the anomalous dimensions $\eta^{xy}_{\text{JQ}}$ and $\eta^{z}_{\text{JQ}}$ of the EPJQ model
at the estimated critical point $q_c$ for $\Delta=1/2$. 
The squares of order parameters
are graphed vs $L$
and analyzed with powerl-law fits (lines on the log-log plots). The off-diagonal spin order parameter square $M^{2}_{xy}(L)$ in
(a) and the dimer oreder parameter square $D^{2}(L)$ in (b) give $\eta^{xy}_{\text{JQ}}=0.13(3)$. The
diagonal spin order parameter square $M^{2}_{z}(L)$ in (c) gives $\eta^{z}_{\text{JQ}}=0.91(3)$.}
\label{fig:JQ_anomalous}
\end{center}
\end{figure}

To determine the anomlous dimensions, i.e., the critical correlation-function exponents $\eta^{xy}_\text{JQ}$ and $\eta^{z}_\text{JQ}$ in
\eqnref{eq:JQexponents}, we analyze the system-size dependence of the squares of the easy-plane off-diagonal spin order parameter
$M^{2}_{xy}(L)$ in \eqnref{eq:mxydef} and dimer order parameter
\begin{equation}
D^{2}(L)=\frac{1}{2}\left ( ((D^{x}(L))^{2}+(D^{y}(L))^{2} \right ),
\end{equation}
where the $x$ and $y$ dimer operators are the appropriate Fourier transforms of Eqs.~(\ref{dops}) corresponding to a columnar VBS;
\begin{subequations}
\begin{align}
&D^{x}=\frac{1}{L^2}\sum_{i}(-1)^{x_i}D^{x}_{i},\\
&D^{x}=\frac{1}{L^2}\sum_{i}(-1)^{y_i}D^{y}_{i}.
\end{align}
\label{eq:dx2def}
\end{subequations}
\noindent In the diagonal $S^z$ channel we study the system size dependence of the staggered magnetization
\begin{equation}
M^{2}_{z}=\frac{1}{L^4}\sum_{i,j}(-1)^{i+j}S^{z}_{i}S^{z}_{j}.
\label{eq:mzdef}
\end{equation}
All these integrated correlation functions should scale as the correlation functions in \eqnref{eq:JQexponents} with the distance $|{\bf r}_{ij}|$
replaced by the system length $L$.

This dimer order parameter should be governed by the same exponent $\eta^{xy}_\text{JQ}$ as the off-diagonal spin order parameter if the predicted
O(4) symmetry is manifested. In contrast, the diagonal magnetic order parameter is associated with a different (larger) anomalous dimension
$\eta^{z}_{\text{JQ}}$, according to the table in \figref{fig:phasediagrams}. We have evaluated the order parameters at $q=0.620$, consistent with
the value of $q_c$ determined in previous subsection. Results are shown in \figref{fig:JQ_anomalous} for system sizes up to $L=32$ and $L=40$ for
the dimer and spin order parameters, respectively. In \figref{fig:JQ_anomalous}(a,b) we show that $M^2_{xy}$ and $D^2$ can be fitted with the same
exponent, $\eta^{xy}_{\text{JQ}}=0.13(3)$, while the fit to $M^2_{z}$ in \figref{fig:JQ_anomalous}(c) delivers a distinctively different exponent;
$\eta^{z}_{\text{JQ}}=0.91(3)$.

\section{Results for the Bilayer Honeycomb model}
\label{sec:numericalbh}

In this section we present our numerical results on the continuous phase transition in the BH model, where an interaction-driven phase transition
between a BSPT phase and a trivial Mott insulator is investigated via large-scale DQMC simulations \cite{mengQSH2,He2016Topinva,He2016Topinvb}
in the ground-state projector version\cite{AssaadEvertz2008}. Acting with the  operator ${\rm e}^{-\Theta H}$ on a trial state (a Slater determinant)
with the projection 'time' $\Theta$ large enough for converging the finite system to its ground state, we simulated linear system sizes $L=12,15,18,21$
and $24$, with  $\Theta=50$ for $L\le 18$, $\Theta=55$ for $L=21$, and $\Theta=60$ for $L=24$. The imaginary-time discretization step was
$\Delta\tau =0.05$, which is sufficiently small to not lead to any significant deviations of scaling behaviors from the $\Delta\tau=0$ limit.

\subsection{The continuous topological phase transition}
\label{sec:BHderivatives}

\begin{figure}[t]
\begin{center}
\includegraphics[width=80mm]{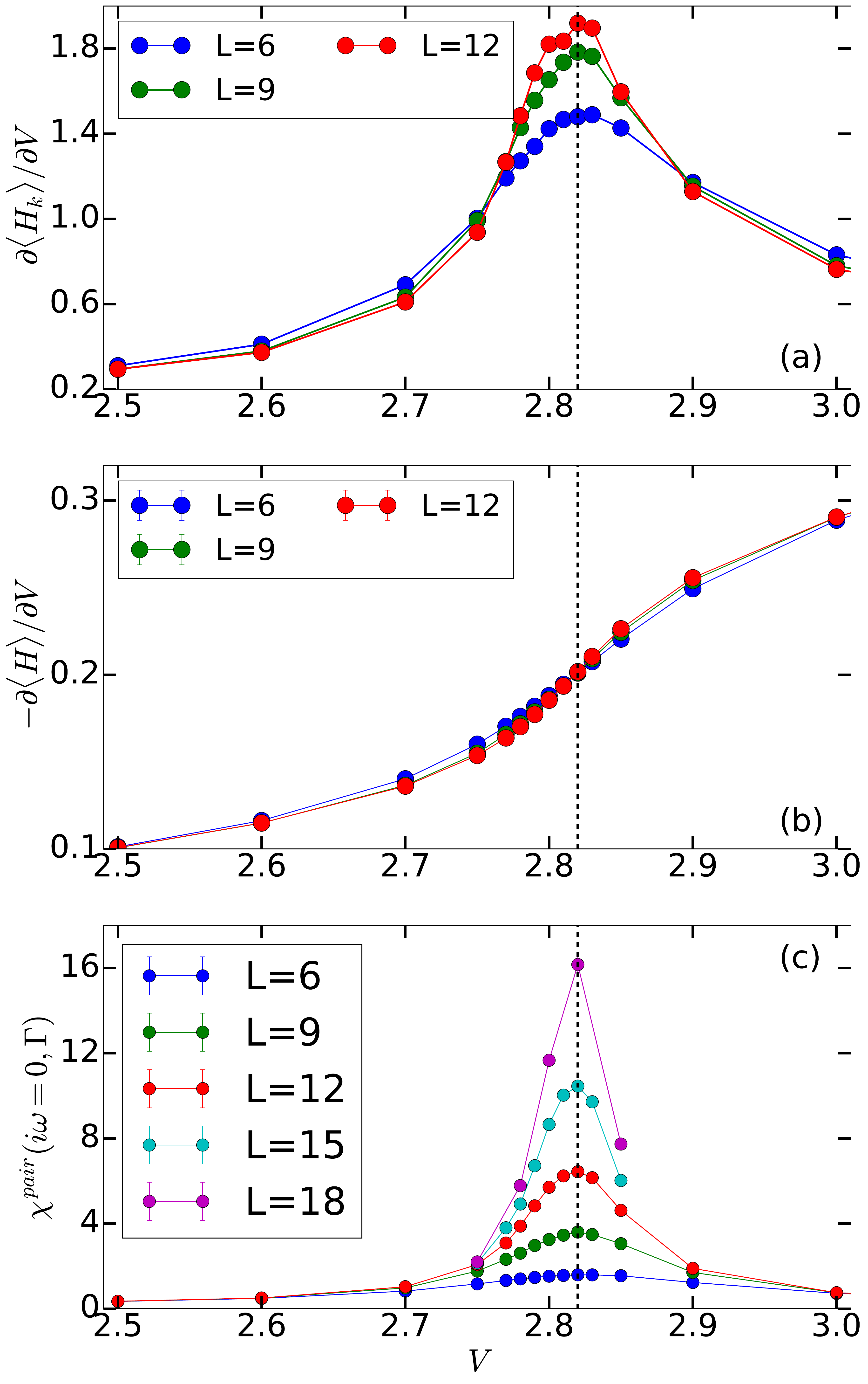}
\caption{DQMC results for the BH model close to its phase transition. The derivative with respect to the coupling $V$ of (a) the kinetic energy
density and (b) the total ground-state energy density for linear system sizes $L=6$, $9$ and $12$. Panel (c) shows the zero-frequency susceptibility of
the O($4$) vector in the pairing channel for $L=6$, $9$, $12$, $15$, $18$. The dashed line indicates our estimated  critical point $V_c=2.82(1)$.}
\label{fig:BH_energyderivative}
\end{center}
\end{figure}

We first present simulation results supporting a continuous BSPT--Mott transition (as was also previously discussed in Ref.~\cite{mengQSH2,He2016Topinvb}).
Figure \ref{fig:BH_energyderivative}(a) shows the derivative of the kinetic energy density of the BH Hamiltonian \eqnref{eq:BH} with respect to the
control parameter $V$ of the phase transition,
\begin{equation}
\frac{\partial \langle H_{\text{k}}\rangle}{\partial V} =
-\frac{t}{N}\frac{\partial}{\partial V}\sum_{\langle ij\rangle}\langle c^{\dagger}_{i}c_j+\text{h.c.} \rangle.
\end{equation}
Here a broad peak develops close to $V_c$, but there is no sign of a divergence, as would be expected at a first-order transition.
Figure \ref{fig:BH_energyderivative}(b) shows the derivative of the ground state energy density, which can be conveniently evaluated by
invoking the Hellmann-Feynman theorem;
\begin{equation}
\frac{\partial \langle H \rangle}{\partial V}=\Big\langle\frac{\partial H}{\partial V}\Big\rangle=
\frac{1}{N}\sum_{i}\langle c_{i1\uparrow}^\dagger c_{i2\uparrow} c_{i1\downarrow}^\dagger c_{i2\downarrow}+\text{h.c.}\rangle.
\label{eq:HellmannFeynman}
\end{equation}
For this derivative a first-order transition would lead to a sharp kink developing with increasing $L$ (corresponding to a real or avoided
level crossing). The smooth behavior supports a continuous phase transition, though of course a very weak first-order transition would produce
a visible singular behavior only for larger system sizes than we consider here.

To determine the phase-transition point $V_c$, we have further
calculated the zero-frequency susceptibility of the O(4) vector (where we here take the on-site spin-singlet pairing operator), defined as
\begin{equation}
\chi^{\rm pair}(i\omega=0,\boldsymbol{\Gamma})=\int_{0}^{+\infty}S(\tau,\boldsymbol{\Gamma})d\tau,
\label{susceptibility}
\end{equation}
where the dynamic pair-pair correlation function is defined as
\begin{equation}
\label{eq:pairpair}
S(\tau,\boldsymbol{\Gamma}) = \frac{1}{L^2}\sum_{ij}e^{i\mathbf{k}\cdot(\mathbf{R}_i-\mathbf{R}_j)}\frac{\langle \Delta_i^{\dagger}(\tau)\Delta_j+\Delta_j^{\dagger}(\tau)\Delta_i \rangle}{2}, \\
\end{equation}
where $k=0$ and $\Delta_i$ defined in Eq.~(\ref{eq:O4vector}). As demonstrated in
\figref{fig:BH_energyderivative}(c), this quantity exhibits a sharp peak, as expected at a gapless critical point with power-law correlations
in both space and time. The divergence is considerably slower than the $\propto$ space-time-volume behavior expected at a first-order
transition. Due to the large computational effort needed for these DQMC simulations, we do not have a sufficient density of points close
to $V_c$ to carry out a systematic analysis of the drift of the peak position, but the data nevertheless allow us to roughly estimate the
convergence to the critical point $V_c/t=2.82(1)$.

\begin{figure}[t]
\begin{center}
\includegraphics[width=82mm]{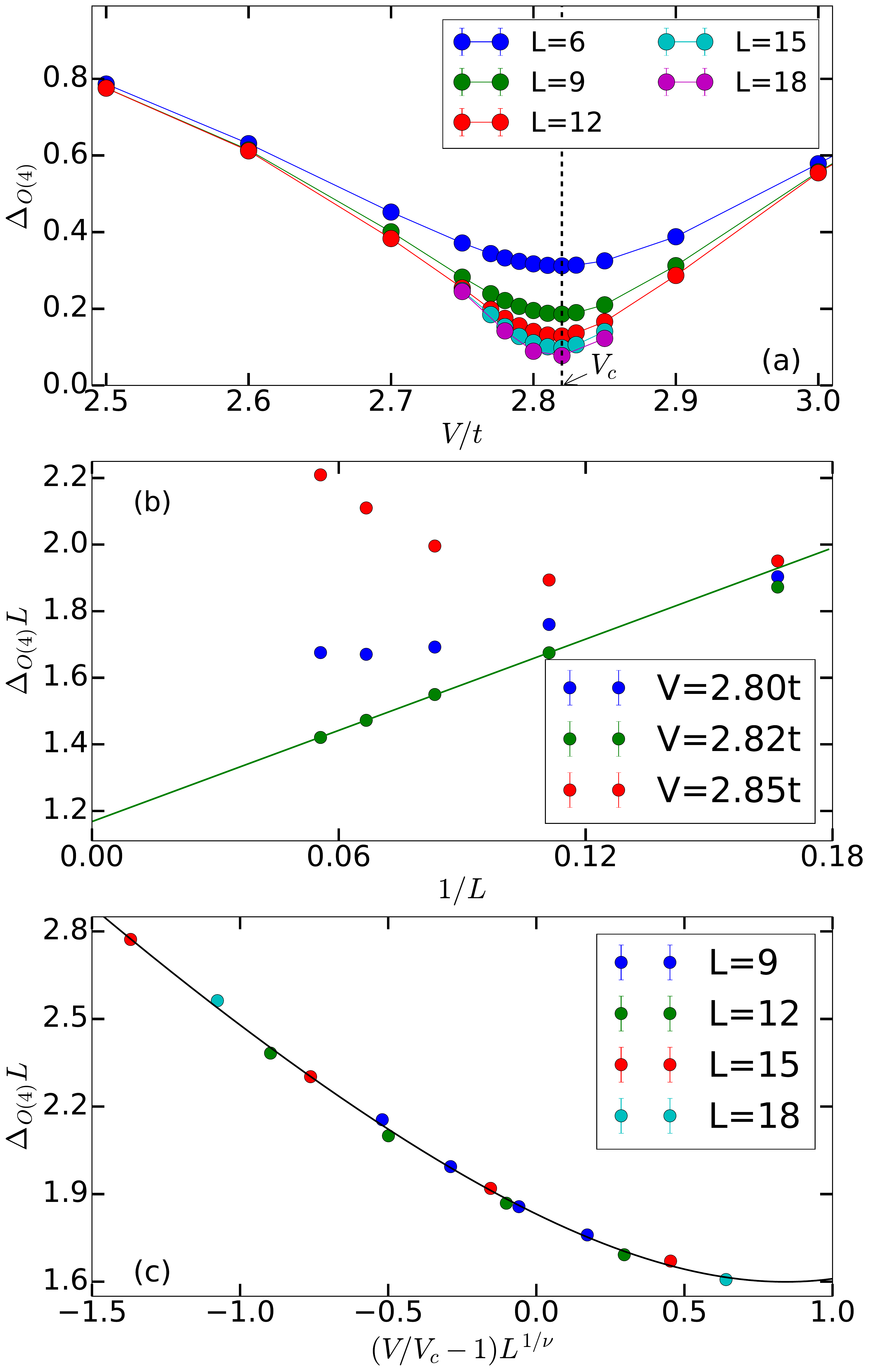}
\caption{(a) Excitation gaps computed from the imaginary-time decay of the O(4) vector across the BSPT-Mott transition for the BH model
with $L=6,9,12,15,18$. A gap closing at $V_c=2.82 t$ (vertical dashed line) is expected. (b) Finite-size scaling of $L\Delta_{O(4)}$ for $V/t=2.80,2.82,2.85$
vs the inverse system size. The behavior at $V/t=2.82$ indicates $z=1$. (c) Data collapse of the data in (a) according to the expected scaling
form Eq.~(\ref{o4scaleform}), yielding $\nu_{\text{BH}}=0.53(5)$ and $V_c/t=2.80(1)$.}
\label{fig:BHO4gap}
\end{center}
\end{figure}

\subsection{O(4) gap and BH correlation-length exponent}
\label{sec:GapAndNu}

We have also extracted the excitation gaps, $\Delta_{O(4)}$, corresponding to the $O(4)$ vectors defined in Eq.~(\ref{eq:O4vector}). According to
Refs.~\cite{mengQSH2,He2016Topinvb} and Eq.~(\ref{eq:pairpair}), the O(4) gap is obtained from the imaginary-time decay of the dynamical O(4) vector
correlation function, and, as discussed in Sec.~\ref{sec:fieldtheories} and Sec.~\ref{sec:BHmodel}, O(4) bosonic modes are expected to become gapless
(with power-law correlation) at the BSPT-Mott critical point. Results for $\Delta_{O(4)}$ as a function of $V/t$ for system sizes $L=6,9,12,15,18$ close
to $V_c$ are presented in \figref{fig:BHO4gap}(a). As expected, $\Delta_{O(4)}$ from every system size $L$ exhibits a dip close to $V_c$, with the gap
minimum decreasing with $L$ as expected with an emergent gapless point at $V_c$. In \figref{fig:BHO4gap}(b) we analyze the size dependence of the
gap at three different coupling values; $V/t=2.80,2.82,2.85$. At $V/t=2.82$, $L\Delta_{O(4)}$ extrapolates linearly in $1/L$ to a nonzero value,
showing that the leading behavior of $\Delta_{O(4)}$ at $V_c$ is $1/L$. This is in line with the expectation that the dynamic exponent $z=1$ at
the BSPT-Mott transition (and is required also for the proposed duality). The behaviors of $\Delta_{O(4)}L$ at $V/t=2.80,2.85$ indicate eventual
divergencies when $L \to \infty$, as expected on either side of the quantum critical point. This constitutes strong evidence of a continuous
transition, instead of a first-order transition at which one instead expects a gap closing exponentially fast.

To extract the correlation-length exponent $\nu_{\text{BH}}$, we have performed data collapse with the $O(4)$ gap away from
the critical point, as shown in \figref{fig:BH_energyderivative}(c). Here we focus on the regime $V > V_c$, where we find less scaling
corrections than for $V< V_c$ and an almost perfect data-collapse according to the expected quantum-critical form
\begin{equation}
\Delta_{O(4)}L=f[(V/V_c-1)L^{1/\nu_{\rm BH}}].
\label{o4scaleform}
\end{equation}
Treating both $\nu_{\text{BH}}$ and $V_c$ as free parameters, the best data collapse delivers $\nu_{\rm BH}=0.53(5)$ and $V_c/t=2.80(1)$. This value of
$V_c$ agrees quite well with the result $V_c=2.82(1)$ estimated roughly from the susceptibility peak in Fig.~\ref{fig:BH_energyderivative}, adding
further credence to the analysis of the critical point even with the rather limited range of system sizes accessible (as compared with the EPJQ model).
Moreover, since we already determined $\eta^{z}_{\text{JQ}}=0.91(3)$ in the EPJQ model, we can use the duality relationship in \eqnref{eq:dualityrelations1}
to predict that the correlation-length exponent of the BH model should be $\nu_{\text{BH}}\sim 0.49(2)$, which is fully consistent the number
determined from the $O(4)$ gap.

\subsection{Anomalous dimensions}

Finally, we study the critical equal-time correlations in the BH model. Here we have used $V=2.817$ for the longest simulations. This value
is within the error bars of the critical value $V_c = 2.82(1)$ and, as we will also show, there are no statistically detectable differences between
data at $V=2.820$ and $2.817$ for the quantities studied in this subsection.

\begin{figure}[t]
\begin{center}
\includegraphics[width=80mm]{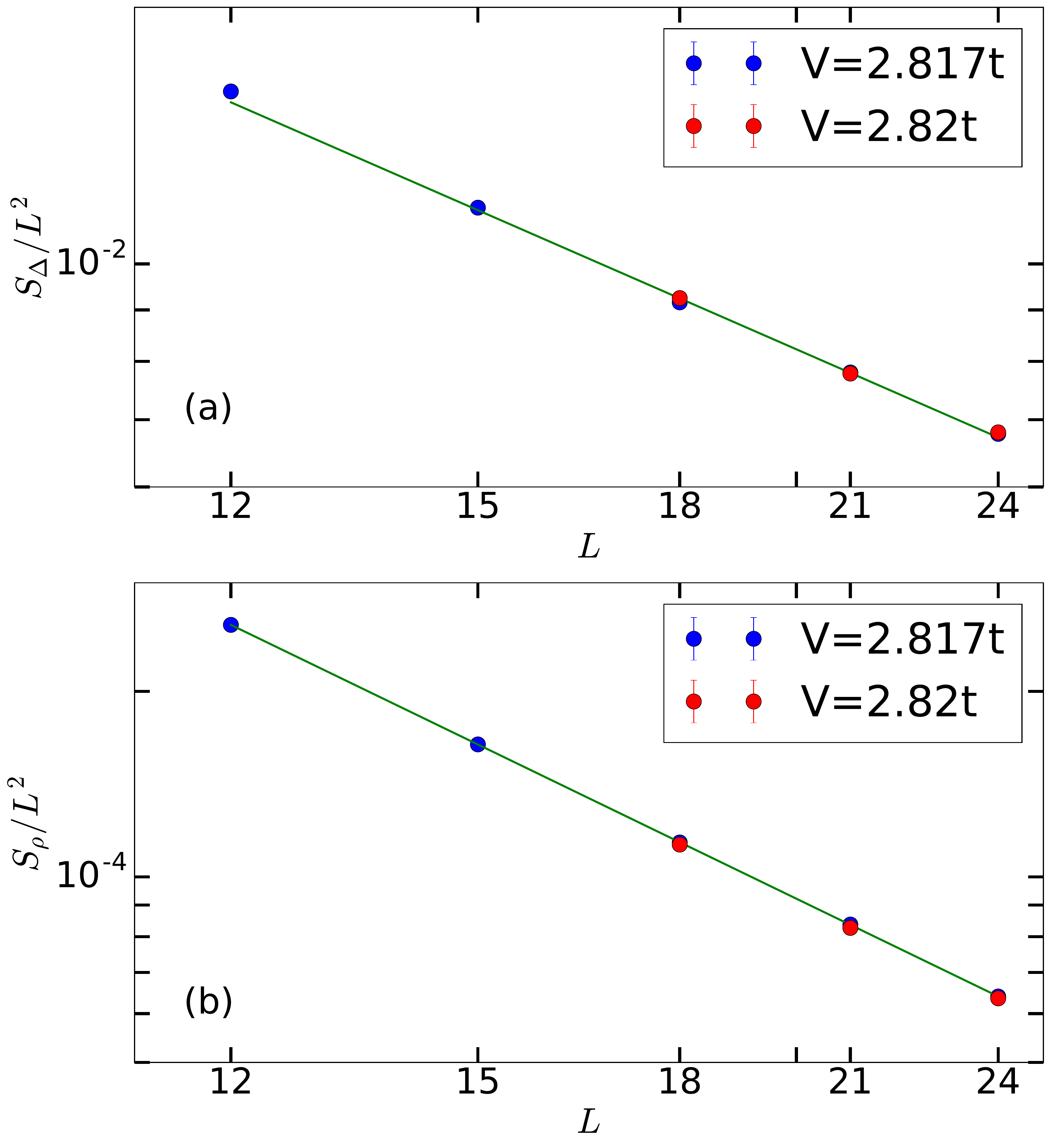}
\caption{Analysis of the anomalous dimensions $\eta^{\Delta}_{BH}$ and $\eta^{\rho}_{\text{BH}}$ of the BH model from squared order parameters
close to the critical point. Only the $V=2.817$ data points for each quantity are used in the fits, and the three points for
$V=2.82$ do not exhibit any deviations from the $V=2.817$ values within the error bars. (a) shows the $O(4)$ order parameter defined in Eq.~(\ref{Delta04squared}).
A power-law fit (straight line on the log-log scale) to the $L \ge 15$ data delivers the exponent $\eta^{\Delta}_{\text{BH}}=0.10(1)$. (b) The pair-density 
order parameter Eq.~(\ref{rho2squared}). Here the power-law fit works well for all system sizes and we obtain $\eta^{\rho}_{\text{BH}}=1.00(1)$.}
\label{fig:BH_anomalous}
\end{center}
\end{figure}

Using one of the components of the $O(4)$ order parameter,
$\langle\Delta^{\dagger}_i\Delta_j\rangle$ with $\Delta_i$ defined in \eqnref{eq:O4vector}, we again construct a squared order parameter.
We can use the susceptibility Eq.~(\ref{eq:pairpair}) to define a corresponding equal-time spatially-integrated correlation function,
\begin{equation}
\frac{S_{\Delta}}{L^2}=\frac{S(\tau=0,{\bf \Gamma})}{L^2},
\label{Delta04squared}
\end{equation}
where the normalization gives the same scaling behaviors as in \eqnref{eq:BHexponents} with the distance $|{\bf r}_{ij}|$ replaced by $L$.
The analysis illustrated in Fig.~\ref{fig:BH_anomalous}(a) indicates a very good power-law scaling, with deviations seen only for the smallest
system size (which we exclude from the fit). The fit delivers the exponent $\eta_{\rm BH}^\Delta=0.10(1)$, which is fully consistent with the
EPJQ exponent  $\eta^{xy}_{\text{JQ}}=0.10(2)$ obtained in Sec.~\ref{sec:JQ_anomalous}. Hence, the duality relation \eqnref{eq:dualityrelations3}
is satisfied to within the statistical errors.

In principle the anomalous dimension can also be obtained from the susceptibility in Fig.~\ref{fig:BHO4gap}(c). Standard scaling arguments
give that the peak height of a generic susceptibility $\chi$ should scale as $\chi_{\rm peak} \propto L^{2-\eta}$ (when the dynamic exponent $z=1$).
We find that the peak in $\chi^{\rm pair}$ scales approximately as $L^2$, i.e., $\eta^{\Delta}$ is very small, but here there appears to be
significant scaling corrections. Moreover, there is large variation of the values for the largest size, $L=18$, close to the critical point,
and we would need additional points to reliably estimate the peak value. We therefore cannot obtain an independent meaningful estimate
for $\eta^{\Delta}_{\rm BH}$ from these data.

We next test the duality relation \eqnref{eq:dualityrelations2}. With $\nu^{xy}_{\text{JQ}}=0.48(3)$ obtained in Sec.~\ref{sec:JQ_crossingpoint} we
expect $\eta^{\rho}_{\text{BH}}\approx 1$. Further, according to Ref.~\cite{Karthik2016}, the exponent $\eta^{\rho}_{\text{BH}}$ should be equal to
the anomalous mass dimension $\eta_\text{QED}$ of $N=2$ QED, for which the value $\eta_\text{QED}=1.0(2)$ was obtained from Monte Carlo simulations.
Thus, we already have good consistency following from the predicted duality between $\eta_\text{QED}$ and $\nu^{xy}_{\text{JQ}}$. Turning to the more
direct test with the BH model, in \figref{fig:BH_anomalous}(b) we plot the squared order parameter corresponding to the pair-density,
\begin{equation}
\label{rho2squared}
\frac{S_{\rho}}{L^2} = \frac{1}{L^4} \sum_{i,j}
\langle \rho_{i\uparrow}\rho_{i\downarrow}\rho_{j\uparrow}\rho_{j\downarrow} \rangle.
\end{equation}
For this quantity, as shown in Fig.~\ref{fig:BH_anomalous}(b), all five system sizes available give results fully consistent with a
power-law decay, with no statistically visible scaling corrections. The fit to the five data points gives the anomalous dimension
$\eta^{\rho}_{\text{BH}}=1.00(1)$, which is consistent with both $\eta_\text{QED}$ and $\nu^{xy}_{\text{JQ}}$, but with a significantly
smaller statistical error. Thus, the duality relation in \eqnref{eq:dualityrelations2} is also confirmed to within error bars.

It is remarkable that the BH model actually seems to give better results (smaller error bars) for the anomalous dimensions than
the EPJQ model, even though system lengths roughly twice as large were used for the latter. The reason is that the statistical errors
on the raw data are smaller in the DQMC simulations. It is still possible that there are scaling corrections present that are not
clearly visible with such a small range of system sizes, and there may then be some corrections to the exponents beyond the purely
statistical error bars (one standard deviation) reported above. The EPJQ results are important in this regard as they seem to show
no significant corrections even with the considerably larger range of system sizes. The lack of significant corrections is also
supported by the good agreement between the non-reivial BH and EPJQ exponents, as predicted by the duality conjecture.

\section{Discussion}
\label{sec:discussion}

We have performed detailed numerical tests on the recently
proposed duality summarized by the predicted exponent
relationships between the BH and EPJQ models in
Eq.~(\ref{eq:dualityrelations}). The relationships were confirmed
within statistical precision at the level of a few percent in the
values of the critical exponents. The thus confirmed duality of
the underlying low-energy quantum field theories,
\eqnref{duality}, is of great importance and interest in
condensed-matter physics, because it relates two seemingly
different quantum phase transitions that have been individually
under intense studies during the past several years: the bosonic
topological phase transition and the easy-plane deconfined quantum
phase transition. The duality was derived using the more basic
dualities between field theories that involve only one flavor of
matter field, and sometimes also a Chern-Simons term of the
dynamical gauge field.

As a consequence of confirming the particular relationship between
critical exponents, our study also provides quantitative evidence
for the underlying basic dualities for theories with one flavor of
matter
field~\cite{son,wangsenthil0,wangsenthil1,wangsenthil2,maxashvin,seiberg1}.
These basic dualities supported by our work represent a
significant step in our understanding of $(2+1)D$ conformal field
theories. They also form the foundation of a large number of other
recently proposed dualities
\cite{xudual,mrossduality,seiberg2,karchtong,mengxu,Karch2017,anomalyseiberg,sonlong,mrosslong}.
Moreover, they lend support to many other dualities that follow
from the same logic and reasoning, such as the duality of Majorana
fermions discussed in Refs.~\cite{maxxu,soseiberg}.

To follow up on our results and insights presented here,
additional numerical investigations are called for to check other
predictions made within these proposed dualities. For example, in
Ref.~\cite{SO5} it was proposed that the Gross-Neveu fixed point
of the $N=2$ QED is dual to the SU(2) version of the NCCP$^1$
model, and also has an emergent SO($5$) symmetry. This symmetry
has recently been discussed within SU(2) deconfined
quantum-criticality as well, and quite convincing results pointing
in this direction were seen in a three-dimensional loop model
\cite{Nahum2015b}. Scaling with the same anomalous dimension for
both spin and dimer correlators had also been observed already
some time ago in the SU(2) $J$-$Q$ model \cite{Sandvik2012}.
Although we have identified the $N=2$ QED as the bosonic
topological phase transition in our bilayer honeycomb lattice
model, we have so far been unable to find the corresponding
Gross-Neveu fixed point. Identifying the additional interactions
that will be required to drive this transition in the BH model is
an important topic for forther research.

Following previous computational studies of deconfined quantum
phase transition with SU($2$) spin-rotation symmetry in $J$-$Q$
models~\cite{Sandvik2007,Lou2009,Shao2016}, we have here
identified a lattice model---the EPJQ model---hosting a continuous
phase transition between the $U(1)$ (planar) N\'{e}el and VBS
states. The fact that this phase transition is continuous is in
itself an important discovery, given that U($1$)
deconfined-quantum criticality had essentially been declared
non-existent, due to unexplained hints of first-order transitions
in some other planar models and what seems like definite proofs in
other cases \cite{Kuklov08,Jonathan2016,Geraedts2017}. Here (as
further discussed in Appendix \ref{app:planarJQ}) we have shown
that the EPJQ model defined in Eq.~(\ref{eq:JQmodel}) can host
first-order or continuous transitions, depending on the degree of
spin-anisotropy parametrized by the Ising coupling $\Delta$ in
Eq.~(\ref{eq:JQmodel}). At $\Delta=1/2$, we find scaling behaviors
with apparently much less influence of scaling corrections than in
the SU($2$) $J$-$Q$ model at its deconfined critical point
\cite{Shao2016}, i.e., the leading correction exponent $\omega$ is
much larger in the EPJQ model. Interestingly, in both cases the
correlation-length exponent is unusually small, close to $1/2$,
while well-studied transitions such as the O($N$) transitioins in
three dimensions have exponents close to $2/3$. Given its small
scaling corrections and likely tricritical point between
$\Delta=1/2$ and $\Delta=1$, the EPJQ model opens doors for future
detailed studies on exotic phase transitions beyond the Landau
paradigm.

\begin{acknowledgments}
The authors thank Chong Wang, Meng Cheng for helpful
discussions. YQQ, YYH, ZYL and ZYM are supported by  the Ministry of
Science and Technology of China under Grant No. 2016YFA0300502,
the National Science Foundation of China under Grant Nos. 91421304, 11421092, 11474356, 11574359, 11674370, and the National Thousand-Young-Talents Program of
China. YQQ would like to thank Boston University for support under its
Condensed Matter Theory Visitors Program. The work of AS is partly supported through the Partner Group program between the Indian Association for the Cultivation of Science (Kolkata) and the Max Planck Institute for the Physics of Complex Systems (Dresden). AWS is supported by the NSF
under Grant No. DMR-1410126 and would also like to thank the Insitute
of Physics, Chinese Academy of Sciences for visitor support. CX is
supported by the David and Lucile Packard Foundation and NSF Grant
No.~DMR-1151208. We thank the following institutions for allocation of CPU
time: the Center for Quantum Simulation Sciences in the Institute of Physics,
Chinese Academy of Sciences; the Physical Laboratory of High Performance
Computing in the Renmin University of China, the Tianhe-1A platform at
the National Supercomputer Center in Tianjin and the Tianhe-2 platform
at the National Supercomputer Center in Guangzhou.
\end{acknowledgments}

\appendix
\section{Crossing-point analysis}
\label{app:crossingpoint}

To determine the critical point and the critical exponents in an unbiased manner, we
adopt the crossing-point analysis applied and tested for 2D Ising and SU(2) $J$-$Q$ models
in Ref.~\cite{Shao2016}. Such analysis can be further traced back to Fisher's "phenomenological
renormalization", which was first numerically tested with transfer-matrix results for the
Ising model in Ref.~\cite{Luck1985}. Ref.~\cite{Shao2016} presented systematic procedures for
a statistically sound application of these techniques with Monte Carlo data. For easy reference we here
summarize how we have adapted the method to the EPJQ model studied in this paper. For the BH model, due to the
much larger computational cost of the DQMC simulations, we do not have data for enough system
sizes to carry out the analysis in this way, and we instead applied other scaling methods in
Sec.~\ref{sec:numericalbh}.

Considering a generic critical point, with $\delta=q-q_c$ defined as the distance to the
critical point $q_c$---for example, here $q$ can be the control parameter $q=Q/(J+Q)$ that
we used for the EPJQ model or it could be $T-T_c$ for a finite-temperature transition. For any
observable $O$, the standard finite-size scaling form is
\beq
O(\delta,L) =
L^{-{\kappa}/{\nu}}f(\delta L^{{1}/{\nu}},\lambda
L^{-\omega}),
\eeq
where we, for the sake of simplicity, only consider one irrelevant field $\lambda$ and the
corresponding subleading exponent $\omega$. At the critical point, one can Taylor expand
the scaling function,
\beq
O(\delta,L)=L^{-{\kappa}/{\nu}}(a_{0}+a_{1}\delta
L^{{1}/{\nu}}+b_{1}L^{-\omega}+\cdots). \label{eq:observable}
\eeq
If one now takes two system sizes, e.g., $L_{1}=L$ and $L_{2}=rL$ $(r>1)$,
and trace the crossing points $\delta^{*}(L)$ of curves $O(\delta,L_1)$ and $O(\delta,L_2)$ versus $\delta$,
one finds
\begin{eqnarray}
\delta^{*}(L)&=& \frac{a_{0}}{a_1}\frac{1-r^{-{\kappa}/{\nu}}}{r^{{(1-\kappa)}/{\nu}-1}}L^{-{1}/{\nu}} \nonumber \\
&& + \frac{b_1}{a_1}\frac{1-r^{-({\kappa}/{\nu}+\omega})}{r^{{(1-\kappa)}/{\nu}-1}}L^{-({1}/{\nu}+\omega)}.
\label{eq:deltafiniteL}
\end{eqnarray}
Now if the quantity $O$ is
asymptotically size-independent (dimensionless) at the critical
point, for example the Binder cumulant (which we write here with a constant and factor
corresponding to a planar vector order parameter),
\beq
U=2\left (1-\frac{\langle m^{4}_{xy}\rangle}{2\langle m^{2}_{xy} \rangle^2}\right)
\eeq
then the corresponding exponent $\kappa=0$, the first term in Eq.~(\ref{eq:deltafiniteL}) with $O=U$ vanishes,
and we obtain the following form for the size-dependent crossing point $\delta^{*}(L)$:
\beq
\delta^{*}(L)=q^{*}_{c}(L)-q_{c}\propto
L^{-({1}/{\nu}+\omega)}, \label{eq:qcshift}
\eeq
hence the shift of the finite-size critical point $q^{*}_{c}(L)$ is approaching the asymptotic value $q_c$ as
$L^{-({1}/{\nu}+\omega)}$.

\begin{figure}[t]
\begin{center}
\includegraphics[width=\columnwidth]{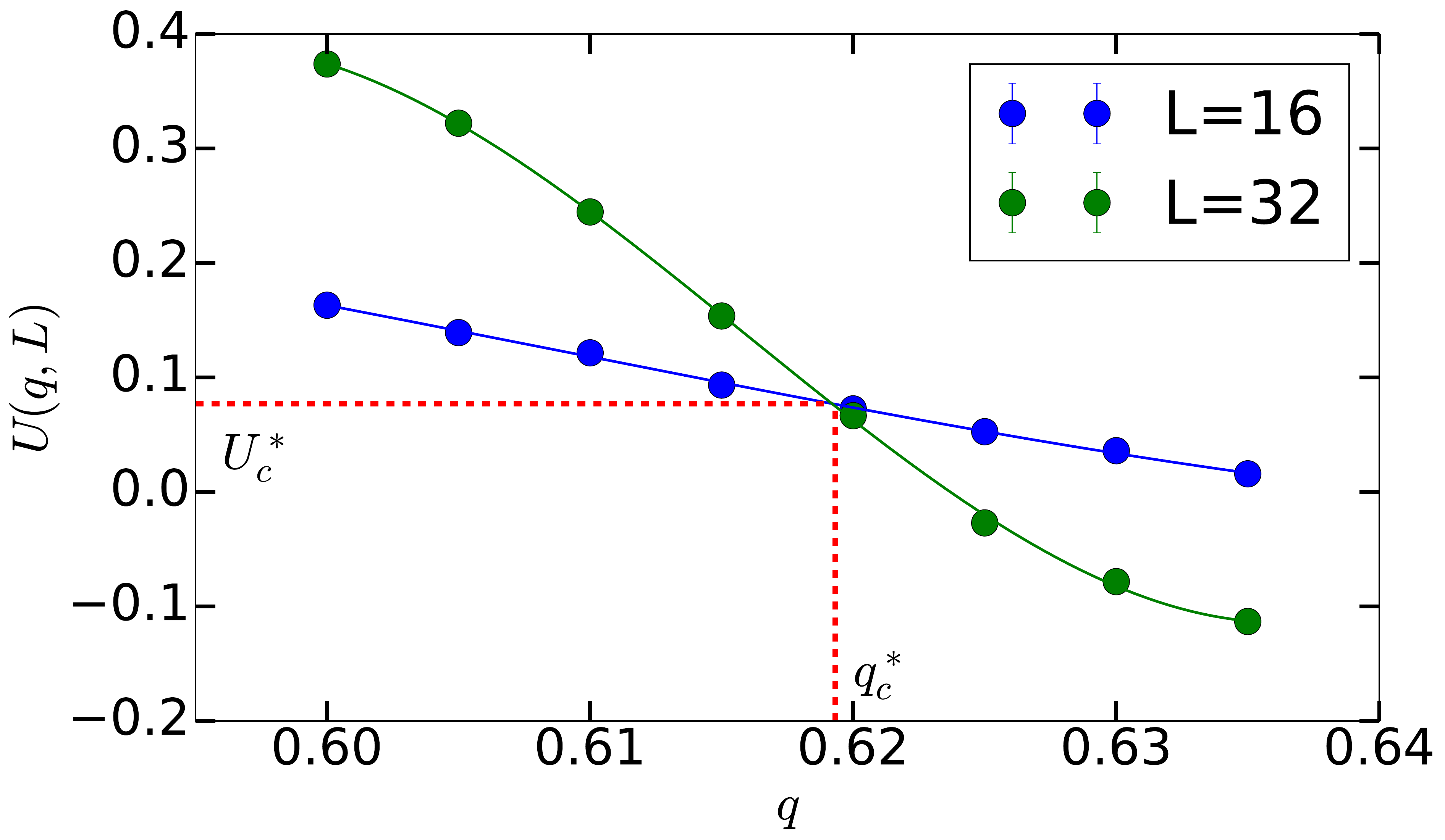}
\caption{Example of finite-size crossing points $q_c^*(L)$,$U_c^*(L)$ of the Binder cumulant,
here for the EPJQ model with $L=16$ and $L=32$. The date are here fitted to third-order
polynomials. The crossing point and derivatives at the crossing point can be determined from the
fitted polynomials.}
\label{fig:s1}
\end{center}
\end{figure}

In practice, one can interpolate within a set of points for each system size by a fitted polynomial, e.g.,
of cubic or quadratic order, and then use these polynomials to find the crossing points. This is
illustrated in Fig~\ref{fig:s1}. Error bars can be obtained by repeating the fits multiple times to data with
Gaussian-distributed noise added. The scaling behavior of $q_c$ predicted by Eq.~(\ref{eq:qcshift}) can be clearly
seen in Fig.~\ref{fig:binder_crossing}(b) of the main text, from which the result ${1}/{\nu}+\omega=4.0(2)$ for
the EPJQ model was obtained.

In addition to obtaining $q^*_c$ and the exponent combination
$1/\nu+\omega$ from the crossing points of the cumulant (or, in principle, some other dimensionless
quantity), one can also use the value $U^*_c$ of the quantity at $q^*_c$, as well as the derivatives
at $q^*_c$, to acquire $\nu$ and $\omega$ independently. We next discuss the derivations underlying these forms.

By inserting $\delta^{*}(L)$ into Eq.~(\ref{eq:observable}), one can obtain the value of observable at the the
finite-size critical point (or, more precisely, the critical point depending on the two sizes, $L$ and $rL$)
$q^{*}_{c}(L)$. It scales as
\beq
O^{*}(L)=L^{-{\kappa}/{\nu}}(a+bL^{-\omega}+\cdots).
\label{eq:Ucshift}
\eeq
Again, for a dimensionless quantity ($\kappa=0$) such as $U$, the deviation of the value at the crossing point from the value in
the thermodynamic limit vanishes with increasing size according to $U^{*}_{c}(L)-U_{c}\propto L^{-\omega}$, an example of which
can be seen in Fig.~\ref{fig:binder_crossing}(c) of the main text---in this case the power-law fit gave the value $\omega=2.3(1)$
of the subleading exponent.

The last step of the analysis of the single dimensionless quantity
is to determine $\nu$ in an independent manner. To this end, one can expand
the quantity $U$ (or any other dimensionless quantity) close to
the critical point,
\beq U(\delta,L)=a_{0}+a_{1}\delta
L^{{1}/{\nu}}+b_{1}L^{-\omega}+c_{1}\delta
L^{{1}/{\nu}-\omega} + \cdots
\eeq
and take the derivative $U'(\delta,L)$ with respect of $\delta$ (in practice with
respect to $q$);
\beq
U'(\delta,L)=a_{1}L^{{1}/{\nu}}+c_{1}L^{{1}/{\nu}-\omega}
+ \cdots.
\eeq
Again, we take two system sizes $L_1=L$ and $L_2=r L$, and at the crossing point
$\delta^{*}(L)$ of the two curves for these system sizes one has
\beqn
U'(\delta^{*},L)&=& a_{1}L^{{1}/{\nu}}+c_{1}L^{{1}/{\nu}-\omega} \nonumber\\
U'(\delta^{*},rL) &=&
a_{1}(rL)^{{1}/{\nu}}+c_{1}(rL)^{{1}/{\nu}-\omega}.
\eeqn
Here we can take the difference of the logarithms of the two equations and obtain,
\beq
\ln \left (\frac{U'(\delta^{*},rL)}{U'(\delta^{*},L)}\right )=\frac{1}{\nu}\ln(r)+dL^{-\omega}
+ \cdots,
\eeq
or, in other words, one can define a finite-size estimate of the  correlation-length exponent $\nu^{*}(L)$
from the finite-size crossing point as,
\beq
\frac{1}{\nu^{*}(L)}=\frac{1}{\ln(r)}\ln\left (\frac{U'(\delta^{*},rL)}{U'(\delta,L)}\right ).
\label{eq:derivativeratio}
\eeq
It can be seen that Eq.~(\ref{eq:derivativeratio}) approaches the thermodynamic
limit correlation-length exponent $\nu$ at the rate ${1}/{\nu^{*}(L)}-{1}/{\nu}=gL^{-\omega}+\cdots$.
This behavior is seen nicely in Fig.~\ref{fig:binder_crossing}(d), where the extrapolation to
infinite size gave $\nu^{xy}_\text{JQ}=0.48(2)$.

\begin{figure}[tp!]
\begin{center}
\includegraphics[width=\columnwidth]{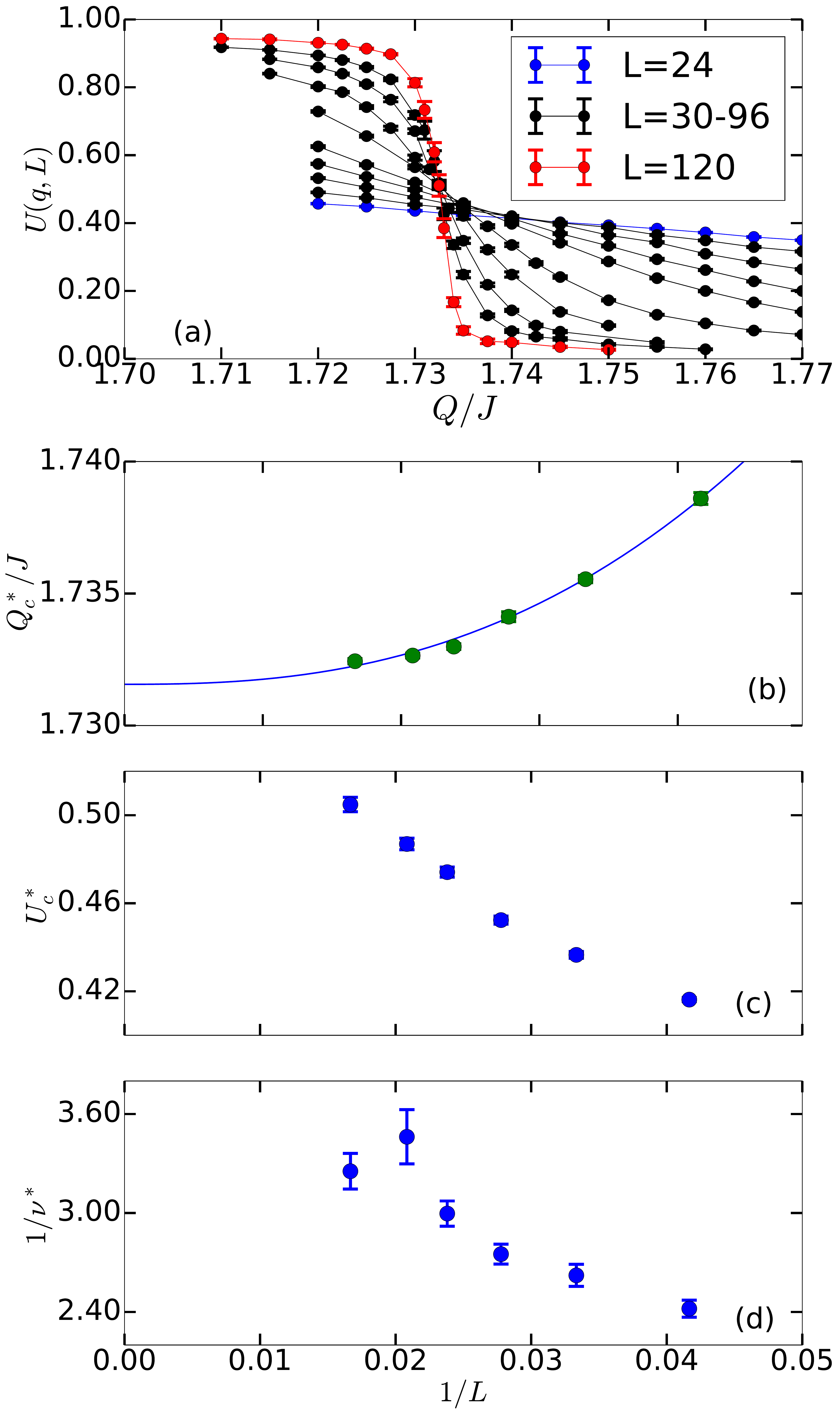}
\caption{Crossing-point analysis of the extreme, $\Delta=1$, version of the EPJQ model, Eq.~~(\ref{eq:planarJQmodel}).
(a) The magnetic Binder cumulant versus the coupling $Q$ in the neighborhood of the phase transition. (b) Size dependence of the
critical point defined as the crossing point of $U$ vs $Q$ for system sizes $L$ and $2L$, with a power-law fit giving $Q_c=1.732(2)$.
(c) The cumulant evaluated at the crossing points. Here it is not clear whether the asymptotic behavior has been reached, and we refrain
from carrying out a fit. (d) The finite-size correlation length. Here as well we do not present any fit, as the asymptotic behavior
is not yet clear.}
\label{fig:planarJQcrossingpoint}
\end{center}
\end{figure}

In principle one can also combine the above method for a dimensionless quantity with some other quantity
$A$, e.g., an order parameter or a long-distance correlation function. Interpolating the data for two system sizes, $A(L)$
and $A(rL)$ at the crossing point of the dimensionless quantity, one can take the logarithm of the ratio and analyze it
in a manner similar to the slope-estimate of $1/\nu$, to yield a series of finite-size estimates for the power-law governing
the size dependence of $A$. This method circumvents the need to know the critical-point value exactly. In practice, since $q_c$
converges fast, it is also appropriate to just use this value and analyze the size-dependence of the quantity $A$ at this
fixed coupling value $q=q_c$, as we did in the main text.

\section{Fully planar EPJQ model}
\label{app:planarJQ}

In the main text we discussed the XYAFM--VBS transition at a fixed anisotropy parameter $\Delta=1/2$ of the EPJQ model.
We here provide some more information on the dependence on $\Delta$.

\begin{figure}[t]
\begin{center}
\includegraphics[width=\columnwidth]{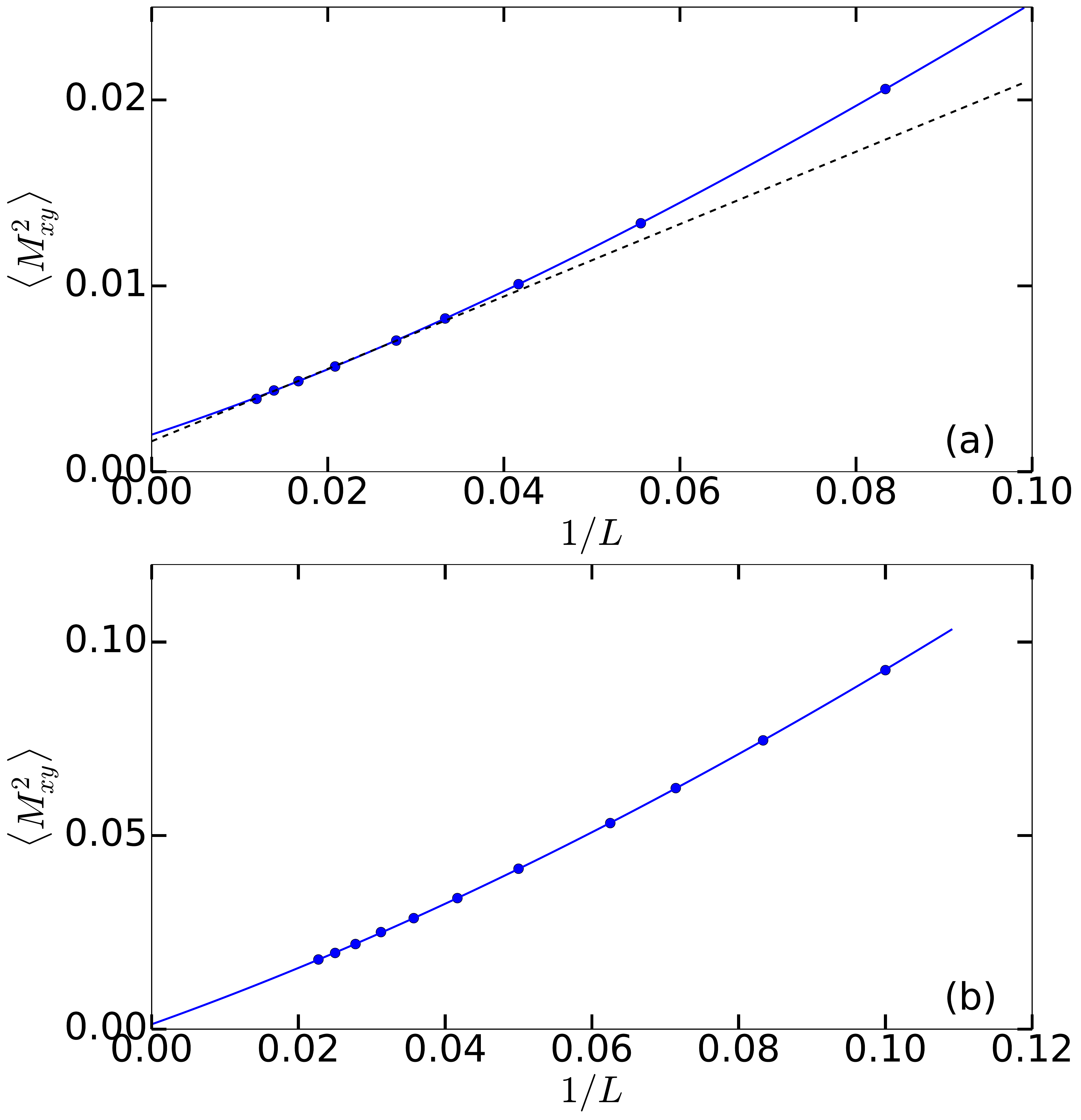}
\caption{Comparison of the XY magnetization squared of the $\Delta=1/2$ and $\Delta=1$ EPJQ models
at their respective critical points. In (a) fits indicating a finite order parameter in the thermodynamic
limit are indicated, with the dashed line corresponding to a purely linear correction and the solid
curve to a quadratic form. In (b) a second-order fit is shown. The extrapolated value at $L=\infty$
is $0$ to within two error bars of the parameter obtained from the fit.}
\label{fig:compareplanarJQ}
\end{center}
\end{figure}

In the extreme planar limit $\Delta \to 1$ of Eq.~(\ref{eq:JQmodel}) we have no $S^{z}S^{z}$ $J$-term and the Hamiltonian is
\begin{equation}
H_\text{JQ}=J\sum_{\langle
ij\rangle}(S^{x}_{i}S^{x}_{j}+S^{y}_{i}S^{y}_{j})+Q\hskip-2mm\sum_{\langle ijklmn\rangle}\hskip-2mm D_{ij}D_{kl}D_{mn}.
\label{eq:planarJQmodel}
\end{equation}
We have analyzed SSE-QMC results for this model in the same way as we did for $\Delta=1/2$ in the main text, using a crossing-point analysis.
The results of this analysis show a distinctively different behavior for $\Delta=1$, pointing to a first-order transition in this case. Results
for the $L$-dependent quantities based on the XY Binder cumulant are shown in Fig.~\ref{fig:planarJQcrossingpoint}. As an aside, we mention
here that for the model in Eq.~(\ref{eq:planarJQmodel}) the off-diagonal spin correlations can be measured as diagonal correlations upon
performing a basis rotation with the $z$ spin components transformed into the $x$ components, and vice versa. Simulating the system with
SSE-QMC in this rotated basis speeds up the simulations of this variant of the model over those for generic $\Delta < 1$. We therefore have
results for larger system sizes in this case.

It is clear from Fig.~\ref{fig:planarJQcrossingpoint} that we can obtain a good estimate of the critical point, but
the cumulant itself and the correlation-length exponent do not exhibit the clear-cut power-law corrections of the type
that we saw for the $\Delta=1/2$ model in Sec.~\ref{sec:numericaljq}. The fact that $1/\nu^{*}$ is larger than $3$ suggests
that the transition may actually be of first order in this case. If so, we would expect the values to eventually tend exactly
to $3$, and the data are consistent with this behavior.

If the transition indeed is of first order, the order parameter should be non-zero at the transition point, reflecting
coexistence of the magnetic and non-magnetic phases. Indeed, as shown in Fig.~\ref{fig:compareplanarJQ}(a), an extrapolation
using a trivial $1/L$ correction, as expected asymptotically for a 2D system breaking a continuous symmetry, indicates a clearly non-zero
value in the thermodynamic limit. The extrapolated value only changes slightly if a higher-order ($1/L^2$) correction
is added (also expected if the order parameter is non-vanishing). In contrast, as shown in Fig.~\ref{fig:compareplanarJQ}(b)
the data for the $\Delta=1/2$ model are fully consistent with no magnetization in the thermodynamic limit. Here the polynomial
fit is strictly not correct, since a non-trivial power is expected at the critical point (which we confirmed in the main
paper), but the extrapolation nevertheless indicates consistency with a vanishing order parameter at the transition in this case.

These results strongly suggest that there is a tricritical point separating a continuous and first-order transition in
the EPJQ model somewhere between $\Delta=1/2$ and $\Delta=1$, which we plan to investigate further in a future study.

\bibliography{dualweb}

\end{document}